\documentclass[a4paper,twocolumn]{revtex4}

\usepackage[english]{babel}
\usepackage{indentfirst}
\usepackage{graphicx}
\usepackage{amsmath}
\usepackage{amssymb}
\usepackage{amsthm}
\usepackage{mathrsfs}
\usepackage{bbold}
\usepackage{lscape}
\usepackage{siunitx}
\usepackage{longtable}
\usepackage{bm}
\usepackage{bbm}
\usepackage{subfigure}
\usepackage{float}
\usepackage{longtable}
\usepackage{color}
\usepackage[utf8x]{inputenc}

\def\ket#1{\left|#1\right>}
\def\bra#1{\left<#1\right|}

\def\bd{b^\dagger}
\def\E#1{\langle #1 \rangle}
\def\eqref#1{Eq.~(\ref{#1})}
\def\figref#1{Fig.~\ref{#1}}

\newcommand{\kreis}[1]{\unitlength1ex\begin{picture}(2.5,2.5)%
\put(0.75,0.75){\circle{3}}\put(0.75,0.75){\makebox(0,0){#1}}\end{picture}}

\begin{document}

\title{Dissipative Phase Transition in Central Spin Systems}
\author{E. M. Kessler$^{1}$, G. Giedke$^{1,5}$, A.
Imamoglu$^2$, S. F. Yelin$^{3,4}$, M. D. Lukin$^4$, J. I. Cirac$^1$}
\affiliation{$^1$ Max-Planck-Institut für Quantenoptik, Hans-Kopfermann-Str. 1
85748 Garching,
  Germany}
   \affiliation{$^2$ Institute of
  Quantum Electronics, ETH-Z\"urich, CH-8093 Z\"urich, Switzerland}
\affiliation{$^3$ Department of Physics, University of Connecticut 2152 Hillside
Road, U-3046 Storrs, CT 06269-3046, USA} \affiliation{$^4$ Department of Physics,
Harvard University, Cambridge, MA 02138, USA}
\affiliation{$^5$ M5, Fakult\"at f\"ur Mathematik, TU M\"unchen, L.-Boltzmannstr. 1, 85748
Garching, Germany}

\vspace{-3.5cm}

\date{\today}
\begin{abstract}
We investigate dissipative phase transitions
in an open central spin system. In our model the central spin interacts
coherently with the surrounding many-particle spin environment and is
subject to coherent driving and dissipation. We develop analytical
tools based on a self-consistent Holstein-Primakoff approximation that
enable us to determine the complete phase diagram associated with the steady
states of this system. It includes first and second-order phase
transitions, as well as regions of bistability, spin squeezing and altered spin pumping dynamics. Prospects of observing these
phenomena in systems such as electron spins in quantum dots or NV
centers coupled to lattice nuclear spins are briefly discussed.
\end{abstract}
 \maketitle

\section{Introduction}
\label{sec:intro}


Statistical Mechanics classifies phases of a given system in thermal equilibrium
according to its physical properties.  It also explains how changes
in the system parameters allow us to transform one phase into
another, sometimes abruptly, which results in the phenomenon of phase
transitions. A special kind of phase transitions occurs at zero
temperature: such transitions are driven by quantum fluctuations instead of
thermal ones
and are responsible for the appearance of exotic quantum
phases in many areas of physics. These quantum phase transitions have been
a subject of intense research in the last thirty years, and are
expected not only to explain interesting behavior of systems at low
temperature, but also to lead to new states of matter with desired properties (e.g., superconductors, -fluids and -solids, topological insulators \cite{Vojta2003,Sachdev:2003fp,Ginzburg:2007dz,Kim:2004hl,Belitz:1994is,Hasan2010}).

Phase transitions can also occur in systems away from their thermal equilibrium. For example, this is the case when  the system interacts with  an environment and, at the same time, is
driven by some external coherent source. Due to dissipation, the environment drives the system to a
steady state, $\rho_0(g)$, which  depends on the system and
environment parameters, $g$. As $g$ is changed, a sudden change in the
system properties may occur, giving rise to a so called {\it
 dissipative phase transition} (DPT)  \cite{HJCarmichael1980,Werner2005,Capriotti2005,Morrison2008b,Eisert:2010vj,Bhaseen12}.   
DPT have been much less studied than
traditional or quantum ones. With the advent of new experimental techniques that allow
to observe them experimentally, they are starting to play an important
role \cite{Baumann2010a}.  Moreover they offer the intriguing possibility of observing
critical effects non-destructively because of the constant intrinsic
exchange between system and environment \cite{Oztop:2011tz}.
In equilibrium statistical mechanics a large variety of toy models exist that describe different kind of transitions. Their study lead to a deep understanding of many of them. In contrast, in the case of DPT few models have been developed.

The textbook example of a DPT occurs in the Dicke model of resonance
fluorescence \cite{Hepp1973,HJCarmichael1980}. There, a system of spins interacts
with a thermal reservoir and is externally driven.
Experimental \cite{Gibbs1976} and theoretical studies \cite{Lawande1981,Puri1980,Bonifacio1978,Schneider2002a} revealed
interesting features such as optical multistability, first and second order phase transitions, and bipartite entanglement.

In this paper, we analyze another prototypical open system:
The model is closely related to the central spin system (CSS) which has been thoroughly studied in
thermal equilibrium \cite{Gaudin1976,BorSt07,SKL03}. In its
simplest form, it consists of a set of spin-1/2 particles (in the following referred to as the \textit{nuclear spins}), uniformly
coupled to a single spin-1/2 (referred to as the \textit{electron spin}). In the model we consider, the central
spin is externally driven and decays through interaction with a Markovian environment.
Recently,
the CSS model has found application in the study of solid
state systems such as electron and nuclear spins in a quantum dot \cite{SKL03} or
an Nitrogen-Vacancy-center.

In what follows we first provide a general framework for analyzing DPT in open systems. In analogy with the analysis of low energy excitations for closed systems, it is based on the study of the excitation gap of the system's Liouville operator $\mathcal L$. 
We illustrate these considerations using the central spin model. 
For a fixed dissipation strength $\gamma$, there are two external parameters one can vary, the Rabi
frequency of the external driving field, $\Omega$, and the Zeemann shift, $\omega$.  We present a
complete phase diagram as a function of those
parameters, characterize all the phases, and analyze the phase
transitions occurring among them.
To this end, we develop a series of analytical tools, based on a
self-consistent Holstein-Primakoff approximation, which allows us to
understand most of the phase diagram. In addition, we use
numerical methods to investigate regions of the diagram where the
theory yields incomplete results. Combining these techniques, we can identify two
different types of phase transitions, and regions of bistability, spin
squeezing, and enhanced spin polarization dynamics. We will also identify regions where anomalous behavior occurs in the
approach to the steady state.
Intriguingly, recent
experiments with quantum dots, in which the central (electronic)
spin is driven by a laser and undergoes spontaneous decay, realize a
situation very close to the one we study here and show effects such as
bistability, enhanced fluctuations, and abrupt changes in
polarization in dependence of the  system parameters \cite{Krebs2010,RFTS07}.

This paper is organized as follows. Section~\ref{sec:frame} sets the
general theoretical framework underlying our study of DPT.  Section~\ref{sec:III} introduces the model, and
contains a structured summary of the main results.
In Section~\ref{sec:PS} we develop the theoretical techniques and use
those techniques to analyze the various phases and classify the
different transitions. Thereafter in Section~\ref{sec:RoB}  numerical techniques are employed to explain the features of the phase diagram which are not captured by the previous theory. Possible experimental realizations and a generalization of the model to inhomogeneous  coupling are discussed in Section~\ref{sec:Implementations}.
 Finally we
summarize the results and discuss potential applications in
Section~\ref{sec:Conclusions}.

\section{General Theoretical Framework}
\label{sec:frame}

The theory of quantum phase transitions in closed systems is a well established and extensively studied area in the field of statistical mechanics.
The typical scenario is the following: a system is described by a Hamiltonian,
$H(g)$, where $g$ denotes a set of systems parameters (like magnetic
fields, interactions strengths, etc). At zero temperature and for a fixed set of parameters, $g$, the system is described by a
quantum state, $\psi_0(g)$, fulfilling $[H(g)- E_{\psi_0}(g)]|\psi_0(g)\rangle=0$,
where $ E_{\psi_0}(g)$ is the ground state energy.
As long as the Hamiltonian is gapped (i.e., the difference between $E_0(g)$ and the first excitation energy is finite), any small change in $g$ will alter the physical properties related to the state $\ket{ \psi_0(g)}$ smoothly and we remain in the same phase. 
However, if the first  excitation gap $\Delta =E_{\psi_1}(g) - E_{\psi_0}(g)$ closes at a given value of the parameters, $g=g_0$, it may happen that the properties change abruptly, in which case a phase transition occurs.  

\begin{centering}
\begin{table*}
\renewcommand{\arraystretch}{1.5}
    \begin{tabular}{|c||c|c|c|l}\hline
 & TPT & QPT & DPT \\
 \hline \hline
  System& Hamiltonian &Hamiltonian  & Liouvillian    \\
operator & $H=H^\dagger$ &$H=H^\dagger$ & $\mathcal L$ -- Lindblad \\
 \hline
Relevant& Free energy & Energy eigenvalues  & "Complex energy" eigenvalues    \\
quantity & $F(\rho)=\E{H}_\rho-T\E{S}_\rho$     &$ E_{\psi}:  H\ket \psi = E_{\psi} \ket \psi$  & $ \lambda_\rho: \mathcal L \rho = \lambda_\rho \rho$\\
 \hline
            & Gibbs state    &Ground state& Steady state \\
 State   & $\rho_T=\underset{\rho\geq0, \textrm{Tr}(\rho)=1}{\textrm{argmin}}[F_\rho]$     &$\ket {\psi_0}=\underset{\|\psi\|=1}{\textrm{argmin}}[\bra \psi H \ket\psi]$&  $\rho_0=\underset{\|\rho\|_\textrm{tr}=1}{\textrm{argmin}}[\|\mathcal L \rho\|_{\textrm{tr}}]$\\
            &$\rho_T\propto \textrm{exp}[-H/k_B T]$    &$[H-E_{\psi_0}]|\psi_0\rangle=0$  & $ {\mathcal L}\rho_0=0$\\
 \hline
 Phase transition& Non-analyticity in $F(\rho_T)$&$\Delta = E_{\psi_1}- E_{\psi_0}$ vanishes&$\textrm{ADR}=\textrm{max}[\textrm{Re}(\lambda_{\rho})]$ vanishes\\
\hline
    \end{tabular}
    \caption{Non-exhaustive comparison of thermal phase transitions (TPT), QPT and DPT. The concepts for DPT parallel in many respects the considerations for QPT and TPT. $||\cdot||_{\textrm{tr}}$ denotes the trace norm and $S$ the entropy. Note that if the steady state is not unique, additional steady states may come with a non-zero imaginary part of the eigenvalue and then appear in pairs: ${\mathcal L}\rho=\pm i y \rho $ ($y\in \mathbb R$).} 
    \label{tab:1}
\end{table*}
\end{centering}

In the following we adapt analogous notions to the case of DPT and introduce the concepts required for the subsequent study of a particular example of a generic DPT in a central spin model. 

We consider a Markovian open system, whose evolution is governed by a time-independent master equation $\dot\rho=\mathcal L(g)\rho$. 
The dynamics describing the system are contractive implying the existence of a steady state. 
This steady state $\rho_0(g)$ is a zero eigenvector to the Liouville superoperator ${\mathcal L}(g)\rho_0(g)=0$. 
This way of thinking parallels that of quantum phase transitions, if one replaces $[H(g)- E_{\psi_0}(g)]\to{\cal L}(g)$. 
Despite the fact that these mathematical objects are very different (the first is a Hermitian operator, and the second a hermiticity-preserving superoperator), one can draw certain similarities between them. 
For instance, for an abrupt change of $\rho_0(g)$ (and thus of certain system observables) it is necessary that the gap in the (in general complex) excitation spectrum of the system's Liouville operator ${\mathcal L}(g)$ closes. 
The relevant gap in this context is determined by the eigenvalue with largest real part different from zero (it can be shown that $\textrm{Re}(\lambda)\leq0$ for all eigenvalues of $\mathcal{L}$ \cite{Rivas:2012jr}).  
The vanishing of the real part of this eigenvalue -- from here on referred to as \textit{asymptotic decay rate} (ADR) \cite{Horstmann:oyrhcBUx} -- indicates the possibility of a non-analytical change in the steady state and thus is a necessary condition for a phase transition to occur. 

In our model system, the Liouvillian low excitation spectrum, and the ADR in particular, can in large parts of the phase diagram be understood from the complex energies of a stable Gaussian mode of the nuclear field. 
We find first order transitions where the eigenvalue of this stable mode crosses the eigenvalue of a metastable mode  at zero in the projection onto the real axis. The real part of the Liouvillian spectrum closes directly as the stable mode turns metastable and vice versa. A finite difference in the imaginary parts of the eigenvalues across the transition prevents a mixing of the two modes and the emergence of critical phenomena such as a change in the nature of the correlations in the steady state at the critical point.
In contrast, we also find a second order phase transition where the ADR vanishes asymptotically as both mode energies become degenerate (at zero) in the thermodynamic limit. Mixing of the two modes at the critical point gives rise to diverging correlations in the nuclear system. 
This observation parallels the classification of quantum phase transitions in closed systems. There, a direct crossing of the ground and first excited state energy for finite systems (mostly arising from a symmetry in the system) typically gives rise to a first order phase transition. An asymptotical closing of the first excitation gap of the Hamiltonian in the thermodynamic limit represents the generic case of a second order transition \cite{Sachdev2011}.

Besides the analogies described so far [cf. Table~\ref{tab:1}], there are obvious differences, like the fact that in DTP $\rho_0(g)$ may be pure or mixed, and that some of the characteristic behavior of a phase may also be reflected in how the steady state is approached.
Non-analyticities in the higher excitation spectrum of the Liouvillian are associated to such dynamical phases.

\section{ Model and Phase Diagram}
\label{sec:III}
\subsection{The Model}
\label{sec:Model}

We investigate the steady state properties of a homogeneous central spin model. The central spin -- also referred to as electronic spin in the following -- is driven resonantly via suitable optical or magnetic fields. Dissipation causes electronic spin transitions from the spin-up to the spin-down state. It 
can be introduced via standard optical pumping techniques.  \cite{Atature:2006ha,Tamarat:2008ko}. Furthermore, the central spin is assumed to interact with an ensemble of ancilla spins -- also referred to as nuclear spins in view of the mentioned implementations \cite{SKL03}-- by an isotropic and homogeneous Heisenberg interaction. In general this hyperfine interaction is assumed to be detuned. Weak nuclear magnetic dipole-dipole interactions are neglected.

After a suitable transformation which renders the Hamiltonian time-independent, the system under consideration is governed by the master equation
\begin{align}
\label{eq:meq}
        \dot{\rho}&=\mathcal{L} \rho\\\nonumber
        &=J \gamma(S^-\rho S^+ - \frac{1}{2} \{ S^+S^-,\rho \}) - i [H_S + H_I + H_{SI},\rho],
\end{align}
where $\{\cdot,\cdot\}$ denotes the anticommutator and
\begin{align}
H_S &= J \Omega (S^+ + S^-),\\
H_I &= \delta\omega I_z , \\\label{eq:hyp}
H_{SI} &= a/2 (S^+I^- + S^- I^+) + a S^+S^- I_z.
\end{align}
$S^{\alpha}$ and $I^{\alpha} = \sum \sigma_i^\alpha$ ($\alpha=+,-,z$) denote electron and collective nuclear spin operators, respectively.
$J \Omega$ is the Rabi frequency of the resonant external driving of the electron (in rotating wave approximation), while $\delta\omega= \omega- a/2$ is the difference of hyperfine detuning $\omega$ and half the individual hyperfine coupling strength $a$. $\delta\omega$ for instance can be tuned via a static magnetic fields in $z$ direction. 
Note that $H_I + H_{SI} = a \vec S\vec I + \omega I_z$, describing the isotropic hyperfine interaction and its detuning. 
The rescaling of the electron driving and dissipation in terms of the total (nuclear) spin quantum number $J$ 
\footnote{Note that the total spin quantum number $J$ is conserved under the action of $\mathcal L$.}
 is introduced here for convenience and will be justified later.
Potential detunings of the electron driving -- corresponding to a term $\Delta S_z$ in the Hamiltonian part of the master equation -- can be neglected if $\Delta \ll Ja$.

In the limit of strong dissipation $\gamma \gg a$ the electron degrees of freedom can be eliminated and \eqref{eq:meq} reduces to
\begin{align}
\label{eq:elim}
\dot\sigma := Tr_S(\dot\rho) = &\gamma_{\textrm{eff}} (I^-\sigma I^+ - \frac{1}{2} \{I^+I^-,\sigma \})\\\nonumber
-& i \left[ \Omega_{\textrm{eff}} I_y + \delta\omega I_z   \right],
\end{align}
where $\gamma_{\textrm{eff}}=\frac{a^2}{\gamma}$, $\Omega_{\textrm{eff}}=\frac{\Omega a}{2\gamma}$ and $\sigma$ is the reduced density matrix of the nuclear system.
This is a generalization of the Dicke model of resonance fluorescence as discussed in \cite{HJCarmichael1980,Schneider2002a,Morrison2008b}.

Master \eqref{eq:meq} has been theoretically shown to display cooperative nuclear effects such as superradiance even for inhomogeneous electron nuclear coupling \cite{Kessler:2010fb}. In analogy to the field of cooperative resonance fluorescence, the system's rich steady state behavior comprises various critical effects such as first and second order DPT and bistabilities. In the following we provide a qualitative summary of the phase diagram and of the techniques developed to study the various phases and transitions.

\begin{figure}[ht]
\centering
\includegraphics[width=0.45\textwidth]{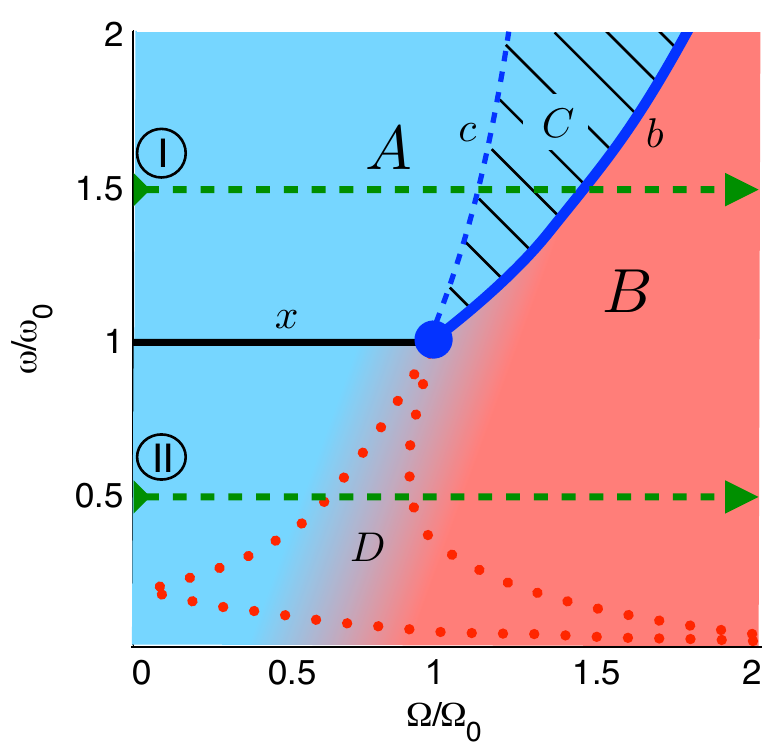}
  \caption{Schematic of the different phases and transitions of Master \eqref{eq:meq}. 
 In the two main phases of the system $A$ (yellow) and $B$ (blue) -- which together cover the whole phase diagram -- the system is found in a RSTSS (cf. text). While phase $A$ is characterized by normal spin pumping behavior (large nuclear polarization in the direction of the dissipation) and a low effective temperature, phase $B$ displays anomalous spin pumping behavior (large nuclear polarization in opposing direction to the dissipation) and high temperature. 
 They are separated by the first order phase boundary $b$ which is associated with a region of bistability $C$ (framed by the boundary $c$). Here a second non-Gaussian solution appears, besides the normal spin pumping mode of $A$. The region of bistability $C$ culminates in a second order phase transition at ($\omega_0,\Omega_0$). Below this critical point the system is supercritical and no clear distinction between phases $A$ and $B$ exists. 
  In this region a dynamical phase $D$ emerges, characterized by anomalous behavior in the approach to the steady state.
   For a detailed description of the different phases and transitions see Section~\ref{sec:phen}.}  \label{fig:0}
\end{figure}

\subsection{Phenomenological Description of the Phase Diagram}
\label{sec:phen}

%

For a fixed dissipation rate $\gamma=a$
\footnote{We discuss the phase diagram exemplarily for the case $\gamma=a$. Also for other values of this order the features of the phase diagram prevail qualitatively. For $\gamma\ll a$ the system becomes increasingly noisy.}
  the different phases and transitions of the system are displayed schematically in \figref{fig:0} in dependence on the external driving $\Omega$ and the hyperfine detuning $\omega$. We concentrate our studies on the quadrant $\Omega,\omega>0$ in which all interesting features can be observed. 
In the following, we outline the key features of the phase diagram.

First we consider the system along the line segment $x$ ($ \omega=\omega_0, \Omega\leq \Omega_0$), where
$\Omega_0=\omega_0=a/2$ define a critical driving strength and critical hyperfine detuning, respectively.
Here $H_I$ vanishes and the steady state can be constructed analytically as a zero-entropy factorized state of the electron and nuclear system. The nuclear field builds up to compensate for the external driving -- forcing the electron in its dark state $\ket \downarrow$ -- until the maximal polarization is reached at the critical value $\Omega_0$. Above this point the nuclear system cannot compensate for the driving $\Omega$ anymore and a solution of different nature, featuring finite electron inversion and entropy is found. The point $\Omega_0$ features diverging spin entanglement and will be identified as a second order phase transition.

For the separable density matrix $\rho_0=\ket{\psi} \bra{\psi}$, $\ket{\psi}=\ket\downarrow \otimes \ket{\alpha}$ the only term in Master \eqref{eq:meq} which is not trivially zero is the Hamiltonian term $S^+ (\frac{a}{2}I^- + J \Omega)$.
However, choosing $\ket{\alpha}$ as an approximate eigenstate of the lowering operator $I^-\ket{\alpha} \approx \alpha \ket{\alpha}$ (up to second order in $\epsilon=1/\sqrt J$) with $\alpha = -2J\Omega /a \equiv - J \Omega/\Omega_0$ ($a$ is the individual hyperfine coupling constant) the corresponding term in \eqref{eq:meq} vanishes in the thermodynamic limit.
In Appendix~\ref{app:exponentialstates} we demonstrate that approximate eigenstates $\ket \alpha$ can be constructed as squeezed and displaced vacua in a Holstein-Primakoff \cite{Holstein1940} picture up to a correction of order 1/J. The squeezing of the nuclear state depends uniquely on the displacement such that these states represent a subclass of {\it squeezed coherent atomic states} \cite{Kitagawa:1993di}. 
Remarkably, this solution -- where along the whole segment $x$ the system settles in a separable pure state  -- exists for all values of the dissipation strength $\gamma$. 

In the limit of vanishing driving $\Omega=0$ the steady state trivially is given by the fully polarized state (being the zero eigenstate of the lowering operator), as the model realizes a standard optical spin pumping setting for dynamical nuclear polarization \cite{Urbaszek2012}.
With increasing $\Omega$, the collective nuclear spin is rotated around the $y$-axis on the surface of the Bloch sphere such that the effective Overhauser field in $x$-direction compensates exactly for the external driving field on the electron spin. 
As a consequence along the whole segment $x$ the dissipation forces the electron in its dark state $\ket \downarrow$,
and all electron observables, but also the entropy and some nuclear observables, are independent of $\Omega$.

Furthermore, the steady state displays increased nuclear spin squeezing in $y$-direction (orthogonal to the mean polarization vector) when approaching the critical point. 
A common measure of squeezing is defined via the spin fluctuations orthogonal to the mean polarization of the spin system. 
A state of a spin-$J$ system is called spin squeezed \cite{Kitagawa:1993di}, if there exists a direction $\vec n$, orthogonal to the mean spin polarization $\E{\vec I}$, such that:
\begin{align}\label{eq:spinsqueezing1}
\xi_{\vec n}^2 \equiv 2 \E{\Delta I_{\vec{n}}^2}/ |\E{\vec I}| < 1.
\end{align}
In \cite{Korbicz2005} it was shown that every squeezed state also contains entanglement among the individual constituents.
Moreover, if $\xi_{\vec n}^2<\frac{1}{k}$ then the spin squeezed state contains
$k$-particle entanglement \cite{Pezze2009,Hyllus2012}.
In Appendix~\ref{app:exponentialstates} we show that the squeezing parameter in $y$-direction for an approximate $I^-$ eigenstate $\ket \alpha$ is given as $\xi_{\hat e_y}^2 = \sqrt{1 -\alpha^2/J^2} + \mathcal O(1/J)=\sqrt{1 -(\Omega/\Omega_0)^2} + \mathcal O(1/J)$. 
Note however, that this equation is valid only for $\xi_{\hat e_y}^2 \geq 1/\sqrt J$. For higher squeezing the operator expectation values constituting the term of order $\mathcal O(1/J)$ can attain macroscopic values of order $\sqrt J$.
For $\Omega \lesssim \Omega_0$ we find that the nuclear spins are in a highly squeezed minimum uncertainty state, with $k$-particle entanglement \footnote{As in
Ref.~\cite{Hyllus2012} we call a pure state $\ket{\psi}$ of $N$-qubits
$k$-particle entangled if $\ket{\psi}$ is a product of states
$\ket{\psi_l}$ each acting on at most $k$ qubits and at least one of these does not factorize.
A mixed state is at least $k$-particle entangled if it cannot be written
as a mixture of $l<k$-particle entangled states.}. 
Close to the critical point $k$ becomes of the order of $\sqrt J$ [$\xi_{\hat e_y}^2 = \mathcal O(1/\sqrt J )$] indicating diverging entanglement in the system.
 
Since the lowering operator is bounded ($|| I^- ||\leq J$), at $\Omega = \Omega_0$ where the nuclear field has reached its maximum value, the zero entropy solution constructed above ceases to exist.
For large electron driving, where $\Omega \gg \Omega_0$ sets the dominant energy scale, the dissipation $\gamma$ results in an undirected diffusion in the dressed state picture and 
in the limit $\Omega \rightarrow \infty$ the system's steady state is fully mixed. 
In order to describe the system for driving strength $\Omega > \Omega_0$, in Section~\ref{sec:theory} we develop a perturbative theory designed to efficiently describe a class of steady states where the electron and nuclear spins are largely decoupled and the nuclear system is found in a fully polarized and rotated state with, potentially squeezed, thermal Gaussian fluctuations (also referred to as \textit{rotated squeezed thermal spin states -- RSTSS} or the \textit{Gaussian mode}) It is fully characterized by its mean polarization as well as the spin squeezing and effective temperature $T_\textrm{eff}$ of the fluctuations (cf. Appendix~\ref{app:RSTSS}). 
Squeezed coherent atomic states, which constitute the solution along segment $x$, appear as a limiting case of this class for zero temperature $T_\textrm{eff}=0$.


We conduct a systematic expansion of the system's Liouville operator in orders of the system size $1/\sqrt{J}$, by approximating nuclear operators by their semiclassical values and incorporating bosonic fluctuations up to second order in an Holstein-Primakoff picture. 
The resulting separation of timescales between electron and nuclear dynamics is exploited in a formalized adiabatic elimination of the electron degrees of freedom. 
The semiclassical displacements (i.e., the electron and nuclear direction of polarization) are found self-consistently by imposing first order stability of the nuclear fluctuations. 
For a given set of semiclassical solutions we derive a second order reduced master equation for the nuclear fluctuations which, in the thermodynamic limit, contains all information on the nuclear state's stability, its steady state quantum fluctuations and entanglement as well as the low excitation dynamics in the vicinity of the steady state and thus allows for a detailed classification of the different phases and transitions.

Using this formalism, we find that the system enters a new phase at the critical point $\Omega_0$, in which the nuclear field can no longer compensate for the external driving, leading to a finite electron inversion and a nuclear state of rising temperature for increasing driving strength.
At the transition between the two phases, the properties of the steady state change non-analytically and
in Section~\ref{sec:wtl} we will find an asymptotic closing of the Liouvillian gap (cf. Section~\ref{sec:frame}) at the critical point, as the Liouvillian's spectrum becomes continuous in the thermodynamic limit.
We will characterize the critical point ($\omega_0, \Omega_0$) as a second order
phase transition. 

Allowing for arbitrary hyperfine detunings $\omega$, a phase boundary emerges from the second order critical point (line $b$ in \figref{fig:0}), separating two distinct phases $A$ (blue) and $B$ (red) of the Gaussian mode. The subregion $C$ of $A$ indicates a region of bistability associated to the phase boundary $b$ and is discussed below.

At $\Omega=0$ the semiclassical equations of motion feature two steady state solutions. Not only the trivial steady state of the spin pumping dynamics -- the fully polarized state in $-z$ direction -- but also an inverted state where the nuclear system is fully polarized in $+z$ direction is a (unstable) solution of the semiclassical system. Quantum fluctuations account for the decay of the latter solution of anomalous spin pumping behavior. 
The two semiclassical solutions (the corresponding quantum states are from here on referred to as the {\it normal} and {\it anomalous spin pumping mode}, respectively) persist for finite $\Omega$. As we show employing the formalism described above (Section~\ref{sec:phases2}), quantum fluctuations destabilize the mode of anomalous behavior in region $A$ of the phase diagram. The stable Gaussian solution in phase $A$ displays a behavior characterized by the competition of dissipation $\gamma$ and the onsetting driving field $\Omega$. The nuclear state is highly polarized in the direction set by the decay, and the electron spin starts aligning with the increasing external driving field. Furthermore, the normal spin pumping mode of phase $A$ is characterized by a low effective spin temperature.

The analysis of the low excitation spectrum of the Liouvillian (Section~\ref{sec:transitions2}) shows a direct vanishing of the ADR at the phase boundary $b$ between $A$ and $B$, while the imaginary part of the spectrum is gapped at all times. At this boundary, the normal mode of phase $A$ destabilizes while at the same the metastable anomalous mode turns stable defining the second phase $B$. 
The two mode energies are non-degenerate across the transition preventing a mixing of the two modes and the emergence of critical phenomena such as diverging entanglement in the system.
Phase $B$ -- anomalous spin pumping -- is characterized by a large nuclear population inversion, as the nuclear field builds up in opposite direction of the dissipation. At the same time the electron spin counter aligns with the external driving field $\Omega$. In contrast to the normal mode of phase $A$, phase $B$ features large fluctuations (i.e., high effective temperature) in the nuclear state, which increase for high $\Omega$, until at some point the perturbative description in terms of RSTSS breaks down and the system approaches the fully mixed state.
Note that region $A$ also transforms continuously to $B$ via the lower two quadrants of the phase diagram (\figref{fig:0}). In this supercritical region \cite{Clifford:1999uv} no clear distinction of the two phases exist.

To complete the phase diagram, we employ numerical techniques in order to study steady state solutions that go beyond a RSTSS description in Section~\ref{sec:RoB}.
The subregion of $A$ labeled $C$ indicates a region of bistability where a second steady state solution (besides the normal spin pumping Gaussian solution described above) appears, featuring a non-Gaussian character with large fluctuations of order $J$. 
Since this mode cannot be described by the perturbative formalism developed in Section~\ref{sec:PS} (which by construction is only suited for low fluctuations $\ll J$) we use numerical methods to study this mode in Section~\ref{sec:RoB} for finite systems. 
We find that the non-Gaussian mode (in contrast to the Gaussian mode of region
$A$) is polarized in $+z$ direction and features large
fluctuations of the order of $J$. Additionally this solution displays large
electron-nuclear connected correlations $\langle S_i I_j\rangle-\langle
S_i\rangle\langle I_j\rangle$. It emerges from the anomalous spin pumping mode coming from region $B$ and the system shows hysteretic behavior in region $C$ closely related to the phenomenon of optical bistability \cite{Bowden1979}. 

A fourth region is found in the lower half of the phase diagram ($D$).
In contrast to the previous regions, area $D$ has no effects on steady state properties. 
Instead the region is characterized by an anomalous behavior in the low excitation dynamics of the system. The elementary excitations in region $D$ are overdamped. 
Perturbing the system from its steady state, leads to a non-oscillating exponential return. This behavior is discussed at the end of Section~\ref{sec:phases2}, where we study the low excitation spectrum of the Liouvillian in this region within the perturbative approach.

In summary, all the phases and transitions of the system are displayed in \figref{fig:0}. Across the whole phase diagram one solution can be described as a RSTSS -- a largely factorized electron-nuclear state with rotated nuclear polarization and Gaussian fluctuations. Phase $A$ hereby represents a region of normal spin pumping behavior. The system is found in a cold Gaussian state, where the nuclear spins are highly polarized in the direction set by the electron dissipation and the electron spin aligns with the external driving for increasing field strength. 
In contrast, phase $B$ displays anomalous spin pumping behavior. The nuclear system displays population inversion (i.e., a polarization opposing the electron pumping direction) while the electron aligns in opposite direction of the driving field. Furthermore the state becomes increasingly noisy, quantified by a large effective temperature, which results in a fully mixed state in the limit of large driving strength $\Omega\rightarrow\infty$.
Along segment $x$ the state becomes pure and factorizes exactly with a nuclear field that cancels the external driving exactly. The nuclear state can be described using approximate eigenstates of the lowering operator $I^-$ which display diverging squeezing approaching the second order critical point $\Omega_0$. From this critical point a first order phase boundary emerges separating phases A and B. It is associated with a region of bistability (area $C$), where a second solution appears featuring a highly non-Gaussian character. The system shows hysteretic behavior in this region.
Region $D$ is a phase characterized by its dynamical properties. The system shows an overdamping behavior approaching the steady state, which can be inferred from the excitation spectrum of the Liouvillian.

Let us now describe the phases and transitions involving the Gaussian mode in detail.


\section{Perturbative Treatment of the Gaussian Mode}
\label{sec:PS}

As seen in the previous section along the segment $x$ the system settles in a
factorized electronic-nuclear state, where the nuclear system can be described
as a lowering operator eigenstate up to second order in  $\epsilon=J^{-1/2}$. 
Motivated by this result we develop a perturbative theory based on a self-consistent Holstein-Primakoff transformation that enables the description of a class of steady states, which generalizes the squeezed coherent atomic state solution along $x$ to finite thermal fluctuations (RSTSS, Appendix~\ref{app:RSTSS}).  
A solution of this nature can be found across the entire phase diagram and we show that this treatment becomes exact in the thermodynamic limit.


In Section~\ref{sec:ptgm} we discuss this Gaussian mode across the whole phase diagram. Steady state properties of the nuclear fluctuations derived from a reduced second order master equation provide deep insights in the nature of the various phases and transitions. 
Observed effects include criticality in both steady state and low excitation spectrum, spin squeezing and entanglement as well as altered spin pumping dynamics.  
Whenever feasible we compare the perturbative results with exact diagonalization techniques for finite systems and find excellent agreement even for systems of a few hundred spins only.
First in Section~\ref{sec:wtl} we apply the developed theory exemplarily along the segment $x$, to obtain further insights in the associated transition at $\Omega_0$.  
In Section~\ref{sec:phases2} we then give a detailed description of the different phases that emerge in the phase diagram due to the Gaussian mode. Thereafter in Section  \ref{sec:transitions2} we conduct a classification of the different transitions found in the phase diagram.

\subsection{The Theory}
\label{sec:theory}
In this section we develop the perturbative theory to derive an effective second order master equation for the nuclear system in the vicinity of the Gaussian steady state.

For realistic parameters, the Liouville operator $\mathcal L$ of
\eqref{eq:meq} does not feature an obvious hierarchy, that would allow for a perturbative treatment. In order to treat the electron-nuclear interaction as a perturbation, we first have to separate the macroscopic semiclassical part of the nuclear fields. 
To this end we conduct a self-consistent Holstein-Primakoff approximation describing nuclear fluctuations around the semiclassical state up to second order. 

The (exact) Holstein-Primakoff transformation expresses the truncation of the collective nuclear spin operators to a total spin $J$-subspace  in terms of a bosonic mode ($b$ denotes the respective annihilation operator):

\begin{align}
\label{eq:HP}
I^-&= \sqrt{2J-b^\dagger b}~ b \\\nonumber
I_z&=b^\dagger b -J.
\end{align}

In the following we introduce a macroscopic displacement $\sqrt J \beta \in \mathbb C$ ($|\beta|\leq2$) on this bosonic mode to account for a rotation of the mean polarization of the state, expand the operators of \eqref{eq:HP} and accordingly the Liouville operator of equation \eqref{eq:meq} in orders of $\epsilon=1/\sqrt{J}$. 
The resulting hierarchy in the Liouvillian allows for an perturbative treatment of the leading orders and adiabatic elimination of the electron degrees of freedom whose evolution is governed by the fastest timescale in the system. 
The displacement $\beta$ is self-consistently found by demanding first order stability of the solution. The second order of the new effective Liouvillian then provides complete information on second order stability, criticality and steady state properties in the thermodynamic limit. 

The macroscopic displacement of the nuclear mode
\begin{align}
b&\rightarrow b+\sqrt{J} \beta,
\end{align}
allows for an expansion of the nuclear operators [\eqref{eq:HP}] in orders of $\epsilon$
\begin{align}
\label{eq:expansion}
I^-/J=
& \sqrt{k}\sqrt{1-\epsilon \frac{\beta \bd + \beta^* b}{k} - \epsilon^2 \frac{\bd b}{k}}~(\beta + \epsilon b)  \\\nonumber
        =&\sum_i \epsilon^i \mathcal{J}^-_{i} ,
\end{align}
where
\begin{align}
\label{eq:Jp}
\mathcal{J}^-_{0}&=\sqrt{k}\beta,\\\label{eq:J1}
\mathcal{J}^-_{1}&=\frac{1}{2\sqrt{k}} \left[(2k-|\beta|^2)b- \beta ^2 \bd\right],\\\nonumber
\mathcal{J}^-_{2}&= -\left[ \frac{\beta^* b + \beta \bd}{ 2\sqrt{k}}b \right. \\
 &\left.  \hspace{1cm}+ \frac{\sqrt{k}\beta}{8 } \left( [\frac{\beta b^\dagger + \beta^* b}{ k}]^2 + 4 \frac{\bd b}{k}\right) \right],\\\nonumber
& \vdots
\end{align}
and $k=2-|\beta|^2$.
Analogously, one finds
\begin{align}\label{eq:Jz0}
I_z/J&=\sum_{i=0}^{2} \epsilon^i \mathcal{J}^z_{i}, \\ \label{eq:Jz}
\mathcal{J}^z_{0}& = |\beta|^2 - 1,\\ \label{eq:Jz1}
\mathcal{J}^z_{1}& = \beta \bd + \beta^* b,\\
\mathcal{J}^z_{2}& = \bd b.
\end{align}

This expansion is meaningful only if the fluctuations in the bosonic mode $b$ are smaller than $\sqrt{J}$.
Under this condition, any nuclear state is thus fully determined by the state of the bosonic mode $b$  and its displacement $\beta$. 


According to the above expansions Master \eqref{eq:meq} can be written as
\begin{align}
\dot{\rho}/J = \left[\mathcal{L}_0  + \epsilon \mathcal{L}_1 +\epsilon^2 \mathcal{L}_2 + \mathcal O (\epsilon^3)\right]\rho,
 \end{align}
where
\begin{align}
\label{eq:L0}
\mathcal{L}_0 \rho= \gamma(&S^-\rho S^+ - \frac{1}{2} \{ S^+S^-,\rho \}_+) \\\nonumber
- i [&S^+(\Omega + a/2 \mathcal J^-_0) + S^-(\Omega + a/2 \mathcal J^+_0) \\\nonumber
&+ a S^+S^- \mathcal J^z_0,\rho],\\
\mathcal{L}_{1,2} \rho= - i [&a/2 (S^+  \mathcal J^-_{1,2} + S^-\mathcal J^+_{1,2}) \\\nonumber
&+ (a S^+S^-  + \delta\omega) \mathcal J_{1,2}^z,\rho].
 \end{align}
 
The zeroth order superoperator $\mathcal{L}_0$ acts only on the electron degrees of freedom. This separation of timescales between electron and nuclear degrees of freedom implies that for a given semiclassical nuclear field (defined by the displacement $\beta$) the electron settles to a quasi-steady state on a timescale shorter than the nuclear dynamics and can be eliminated adiabatically on a coarse grained timescale.
In the following we determine the effective nuclear evolution in the submanifold of the electronic quasi-steady states of $\mathcal L_0$.

Let $P$ be the projector on the subspace of zero eigenvalues of $\mathcal L_0$, i.e., the zeroth order steady states, and $Q=\mathbb 1-P$. Since $\mathcal L_0$ features a unique steady state, we find $P\rho = \textrm{Tr}_S(\rho) \otimes \rho_{ss}$, where $\textrm{Tr}_S$ denotes the trace over the electronic subspace and $\mathcal L_0 \rho_{ss}=0$. By definition it is $P\mathcal L_0=\mathcal L_0P=0$. After a generalized Schrieffer-Wolff transformation \cite{Kessler:QGYw0mMI}, we derive an effective Liouvillian within the zeroth order steady state subspace in orders of the perturbation
\begin{align}
\label{eq:meqeff}
\mathcal L_{\mathrm{eff}} =& \epsilon P\mathcal L_1 P \\\nonumber
                          &+ \epsilon^2(P\mathcal L_2 P - P \mathcal L_1 Q \mathcal L_0^{-1} Q\mathcal L_1P) + \mathcal O(\epsilon^3).
\end{align}
After tracing out the electron degrees of freedom the dynamics of the nuclear fluctuations $b$ are consequently governed by the reduced master equation
\begin{align}
\label{eq:mu}
\dot\sigma :=  \textrm{Tr}_S(P\dot \rho) =  \textrm{Tr}_S(\mathcal L_{\mathrm{eff}}P\rho).
\end{align}

The first order term in $\epsilon$ of \eqref{eq:meqeff} can be readily calculated
\begin{align}
\label{eq:1}
Tr_s(P\mathcal L_1 P\rho) = -i \left[ \E A_{ss} b +\E {A^\dagger}_{ss} \bd ,\sigma  \right] ,
\end{align}
where $A$ is an electronic operator
\begin{align}
A=&\beta^* (aS^+S^-+\delta \omega) \\\nonumber
&+ \frac{a}{4\sqrt{k}} \left[(2k-|\beta|^2)S^+ - (\beta^*)^2 S^-\right].
\end{align}
$ \E A_{ss}$ denotes the steady state expectation value according to $\mathcal{L}_0$, which depends on the system parameters $\gamma$ and $\Omega$ and on the semiclassical displacement $\beta$ via optical Bloch equations derived from  $\mathcal{L}_0$ as described below. \eqref{eq:1} represents a driving of the nuclear fluctuations to leading order in the effective dynamics. Thus for the steady state to be stable to first order, we demand \begin{align}
\label{eq:beta}
 \E A_{ss}=0.
\end{align}
This equation defines self-consistently the semiclassical nuclear displacement $\beta$ in the steady state in dependence on the system parameters $\gamma, \Omega$ and $\delta\omega$.


The calculation of the second order term of \eqref{eq:meqeff} is more involved and presented in Appendix~\ref{app:2ndorder}. 
We find the effective nuclear master equation to second order \footnote{in \cite{Kessler:QGYw0mMI} it has been shown that this type of master equation is of Lindblad form.}
\begin{align}
\label{eq:effmeq}
\dot \sigma =& 2 R_a \left( b\sigma \bd - \frac{1}{2} \{\bd b,\sigma \} \right)\\\nonumber
+& 2R_b \left( \bd\sigma b - \frac{1}{2} \{b\bd ,\sigma \} \right)\\\nonumber
+& c \left( b\sigma b - \frac{1}{2} \{bb ,\sigma \} \right)\\\nonumber
+& c^* \left( \bd\sigma \bd - \frac{1}{2} \{\bd\bd ,\sigma \} \right)\\\nonumber
- & i \left[ (I_a + I_b+F)\bd  b + (\alpha+B^*) b^2 + (\alpha^*+B) (\bd)^2 ,\sigma\right],
\end{align}
with 
\begin{align}
\label{eq:coeff1}
B=& -\frac{a\beta }{16 \sqrt{k^3} }  \left[ ( 4k + |\beta|^2   )  \E{ S^-}_{ss} + \beta^2  \E{ S^+}_{ss}   \right],\\
\label{eq:coeff11}
F=& -\frac{a}{8 \sqrt{k^3} }   ( 4k + |\beta|^2) \left(\beta \E{S^+}_{ss} + \beta^* \E{S^-}_{ss} \right) \\\nonumber
&+ a (\E{S^+S^-}_{ss} + \delta\omega/a),
\end{align}
and
\begin{align}
\label{eq:coeff2}
R_a&= \int_0^\infty dt~ \mathrm{Re}\left( \E{A^\dagger (t) A(0)}_{ss} \right),\\\nonumber
I_a&= \int_0^\infty dt~ \mathrm{Im}\left( \E{A^\dagger (t) A(0)}_{ss} \right),\\\nonumber
R_b&= \int_0^\infty dt~ \mathrm{Re}\left( \E{A (t) A^\dagger(0)}_{ss} \right),\\\nonumber
I_b&= \int_0^\infty dt~ \mathrm{Im}\left( \E{A (t) A^\dagger(0)}_{ss} \right),\\\nonumber
c&= \int_0^\infty dt~ \E{\left\{ A (t), A(0)\right\} }_{ss} ,\\\nonumber
\alpha &= \frac{1}{2i}\int_0^\infty dt~ \E{\left[ A (t), A(0)\right] }_{ss} .
\end{align}

For a given set of system parameters the coefficients defining the nuclear dynamics [\eqref{eq:coeff1}, \eqref{eq:coeff11} and \eqref{eq:coeff2}] depend only on  the nuclear displacement $\beta$. 
After choosing $\beta$ self-consistently to fulfill \eqref{eq:beta} in order to guarantee first order stability, \eqref{eq:effmeq} contains all information of the nuclear system within the Gaussian picture, such as second order stability as well as purity and squeezing of the nuclear steady state. 
Also it approximates the Liouville operator's low excitation spectrum to leading order and thus contains information on criticality in the system. 
\eqref{eq:effmeq} therefore forms the basis for the subsequent discussion of the RSTSS mode and the corresponding phases and transitions in Section~\ref{sec:PS}.

In order to calculate the coefficients of \eqref{eq:coeff2}, we have to determine integrated electronic autocorrelation functions of the type $\int_0^\infty dt~ \E{S^i(t) S^j(0) }_{ss}$ and $\int_0^\infty dt~ \E{S^i(0) S^j(t) }_{ss}$, where $i,j=+,-,z$. 
The dynamics of single electron operator expectation values are governed by the optical Bloch equations derived from $\mathcal L_0$
\begin{align}
\frac{d}{dt} \E{\Delta \vec{S}}=\mathcal M \E{\Delta \vec{S}},
\end{align}
where $\Delta \vec{S}:= \vec{S}- \E{\vec{S}}_{ss}$ and $ \vec{S} =(S^+,S^-,S_z)^T$ and
\begin{align}
\mathcal M=
\begin{pmatrix}
-(\frac{\gamma}{2} - i a\mathcal L_0^z) & 0 & -2i\tilde\Omega^*\\
0   & -(\frac{\gamma}{2} + i a \mathcal L_0^z) & 2i\tilde\Omega \\
-i\tilde\Omega & i\tilde\Omega^* &-\gamma
\end{pmatrix},
\end{align}
where we defined $\tilde\Omega = \Omega + \frac{a}{2} \sqrt{k} \beta$ and $\mathcal L_0^z$ is given in \eqref{eq:Jz}.
The steady state solutions can readily be evaluated
\begin{align}
\label{eq:Z1}
\E{S^+}_{ss}=2i \frac{\tilde\Omega^* (\gamma+2ia \mathcal L_0^z) }{\gamma^2+ 4a \mathcal L_0^{z2} + 8|\tilde\Omega|^2}, \\
\label{eq:Z2}
\E{S_z}_{ss}= - \frac{1}{2} \frac{\gamma^2 + 4a \mathcal L_0^{z2}}{\gamma^2+ 4a \mathcal L_0^{z2} + 8|\tilde\Omega|^2} .
\end{align}

Defining the correlation matrix $\bold S=\E{\Delta \vec S\Delta \vec S^\dagger}_{ss}$ and $\bold S_t=\E{\Delta \vec S_t\Delta \vec S^\dagger}_{ss}$, the Quantum Regression Theorem \cite{Lax:1963cy} yields the simple result:
\begin{align}
\bold S_t =& e^{\mathcal M t} \bold S,\\
\bold S_t^\dagger =& \E{\Delta \vec S\Delta \vec S_t^\dagger}_{ss}  = \bold S e^{\mathcal M^\dagger t}.
\end{align}
Finally the time integrated autocorrelation functions reduce to the simple expression
\begin{align}
\mathcal F_1 =& \int_0^\infty dt \bold S_t =\int_0^\infty dt e^{\mathcal M t} \bold S = - \mathcal M^{-1} \bold S,\\
\mathcal F_2 =& \int_0^\infty dt \bold S_t^\dagger = \mathcal F_1^\dagger = - \bold S \left(\mathcal M^{-1}\right)^\dagger.
\end{align}
These matrices straightforwardly define the coefficients of the effective master equation of the nuclear fluctuations \eqref{eq:effmeq}. In Appendix~\ref{app:coeff} we provide explicit formulas to calculate the relevant coefficients.

\subsection{Phase Diagram of the Gaussian Mode}
\label{sec:ptgm}

In this Section we use the theory  developed above to study the RSTSS mode across the phase diagram.
As outlined in the previous section we first determine self-consistently possible semi-classical displacements $\beta$, which guarantee first order stability [\eqref{eq:beta}].
For each of these solutions we determine the effective master equation for the nuclear fluctuations \eqref{eq:effmeq}, which in the thermodynamic limit contains all information on the steady state and the low excitation dynamics and we discuss properties like second order stability, criticality as well as purity and squeezing of the nuclear steady state. 
Using this information we provide a complete picture of the various phases and transitions involving the RSTSS solution. 


\subsubsection{Methods and General Features}
\label{sec:methods}

In order to determine the semiclassical displacements $\beta$ which guarantee first order stability,  we show in Appendix~\ref{app:beta} that \eqref{eq:beta} is equivalent to the semiclassical steady state conditions.  
Due to a symmetry in the equation, the steady state displacements appear in pairs $\beta_-$, $\beta_+$.
Any semiclassical displacement $\beta$ can be straightforwardly converted to the mean spin polarizations up to leading order in $\epsilon$ according to \eqref{eq:Jp}, \eqref{eq:Jz}, \eqref{eq:Z1}, and \eqref{eq:Z2}.
In the thermodynamic limit the two sets of steady state expectation values extracted from $\beta_-$ and $\beta_+$ share the symmetry $(\pm \E{S_x}, \E{S_y},\E{S_z},\E{I_x},\pm\E{I_y},\pm\E{I_z})$. In large parts of the phase diagram the solution $\beta_-$ ($\beta_+$) displays high nuclear polarization in the same (opposite) direction as the the electron spin pumping. We define the corresponding quantum states as the normal (anomalous) spin pumping mode.

The two solutions $\beta_\pm$ define two corresponding master equations of the nuclear fluctuations around the respective semiclassical expectation values according to \eqref{eq:effmeq}. 
These master equations are subsequently used to determine second order stability of the nuclear fluctuations and, if the dynamics turn out to be stable, the steady state properties of the nuclear system. We emphasize that the effective Master Eq.~(\ref{eq:effmeq}) not only can be used to determine steady state properties, but also reproduces accurately the low excitation spectrum of the exact Liouvillian. It thus also describes the system dynamics in the vicinity of the steady state (increasingly accurate for large $J$).

From \eqref{eq:effmeq} one readily derives a dynamic equation for the first order bosonic moments
\begin{align}
\label{eq:b}
\dot{\begin{pmatrix}
\E{b}\\
\E{b^\dagger}
\end{pmatrix}}
=\Sigma \begin{pmatrix}
\E{b}\\
\E{b^\dagger}
\end{pmatrix},
\end{align}
with
\begin{align}
\Sigma=&\begin{pmatrix}
-(R_a -R_b) -i \chi  &  -2 i \xi\\
2 i \xi^*               &-(R_a -R_b) + i \chi
\end{pmatrix},\\\label{eq:chi}
\chi=&I_a+I_b+F,\\\label{eq:xi}
\xi=&\alpha^* + B,
\end{align}
where all parameters are functions of the semiclassical displacements $\beta_\pm$. 
This equation of motion -- and thus the corresponding master equation itself -- features a fixed point if the eigenvalues of the matrix $\Sigma$ have negative real part ($\textrm{Re}[\lambda_{1,2}]<0$).
Due to the symmetry between $\beta_+$ and $\beta_-$ one finds that the eigenvalues of the two $\Sigma$ matrices corresponding to $\beta_\pm$ fulfill $\textrm{Re}[\lambda_{1,2}(\beta_+)] = - \textrm{Re}[\lambda_{1,2}(\beta_-)]$ such that across the whole phase diagram only one solution is stable at a time and defines the corresponding phase in the phase diagram.
Note however, that the unstable solution decays at a rate that is second order in $\epsilon$. Preparing the system in this state  consequently leads to slow dynamics, such that this solution exhibits metastability.

\begin{figure}[ht]
\centering
\includegraphics[width=0.5\textwidth]{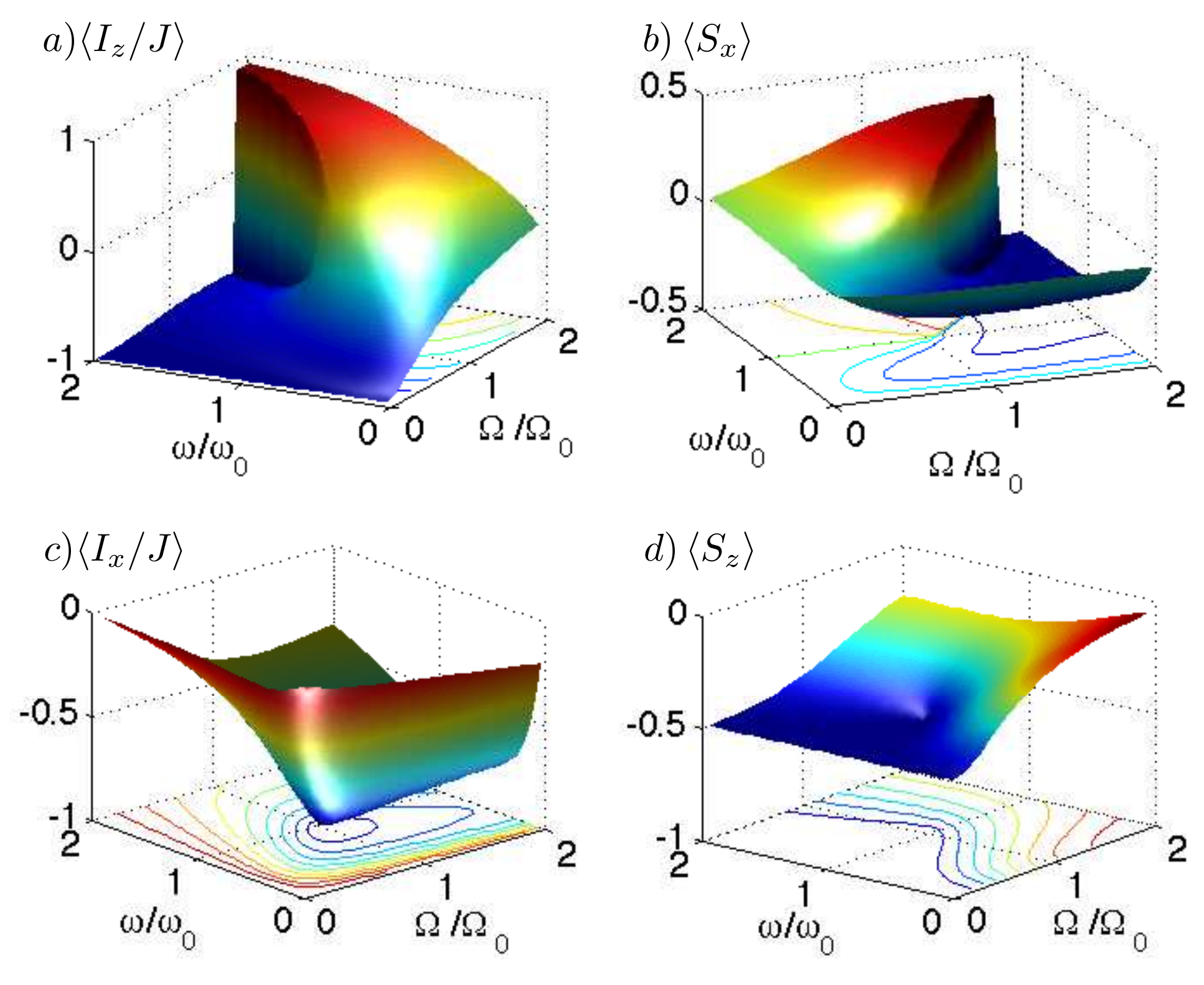}
  \caption{The system observables of the RSTSS solution in the thermodynamic limit show clear signatures of first and second order transitions. 
  (a) The nuclear polarization in $z$-direction $\E{I_z/J}_{ss}$ switches
  abruptly from minus to plus at the phase boundary $b$.  (b) The electron
  polarization in $x$-direction $\E{S_x}_{ss}$ shows a similar discontinuous behavior along $b$. (c)
  The nuclear polarization in $x$-direction changes smoothly across the phase
  boundary $b$. Along the segment $x$ ($\omega=\omega_0, \Omega<\Omega_0$) the
  nuclear field in $x$-direction builds up linearly to cancel the external
  driving. (d) The electron polarization in $z$ direction also does not show
  signatures of the first order transition $b$. Along segment $x$ the electron
  is fully polarized in -$z$ direction up to the second order critical point
  $(\omega_0,\Omega_0)$, where it changes non-analytically (see also \figref{fig:shells}). 
   }  \label{fig:1}
\end{figure}

In the following we implicitly choose the stable $\beta$ for which the real parts of the eigenvalues of $\Sigma$ are negative and discard the unstable solution. 
\figref{fig:1} displays a selection of steady state expectation values in the thermodynamic limit across the phase diagram for the stable solution. Different expectation values illustrate the different nature of  phase $A$ and $B$ and show distinct signatures of first and second order phase transitions  which will be discussed in greater detail in Section~\ref{sec:phases2} and \ref{sec:transitions2}.
The approximate steady state polarizations found in this way coincide with the exact values found via diagonalization techniques 
to an extraordinary degree ($\sim10^{-3}$ relative deviation for J=150). 
Corrections to the perturbative solutions are of the order $1/J$ since the first order expectation values of the bosonic mode vanish by construction, since $\E b =0$ [Compare \eqref{eq:expansion} and \eqref{eq:Jz0}]. In the thermodynamic limit the perturbative solution becomes exact.

The two eigenvalues of $\Sigma$ are typically of the form $\lambda_{1,2}=a \pm i b \ $ (except in region $D$ which will be discussed below) and define the complex energy of the mode.
In this case the matrix $\Sigma$ contains all information on the low excitation spectrum of the Liouvillian which is approximated by multiples of the mode energies within the perturbative treatment
\footnote{The inset of \figref{fig:4} clearly shows this bosonic characteristics of the exact spectrum for J=150. Outside the region of bistability the real part of the spectrum is approximately equidistant. }.
The low excitation spectrum contains information about criticality of the system and the dynamics in the vicinity of the steady state, and will be used to discuss and classify the different transitions in the phase diagram. In particular the eigenvalue of $\Sigma$ with largest real part approximates the ADR in the thermodynamic limit in those regions of the phase diagram where the Gaussian mode is responsible for the lowest excitations in the Liouvillian spectrum (only in the region of bistability $C$ this is not the case).

\begin{figure}[h]
  \centering
\includegraphics[width=0.5\textwidth]{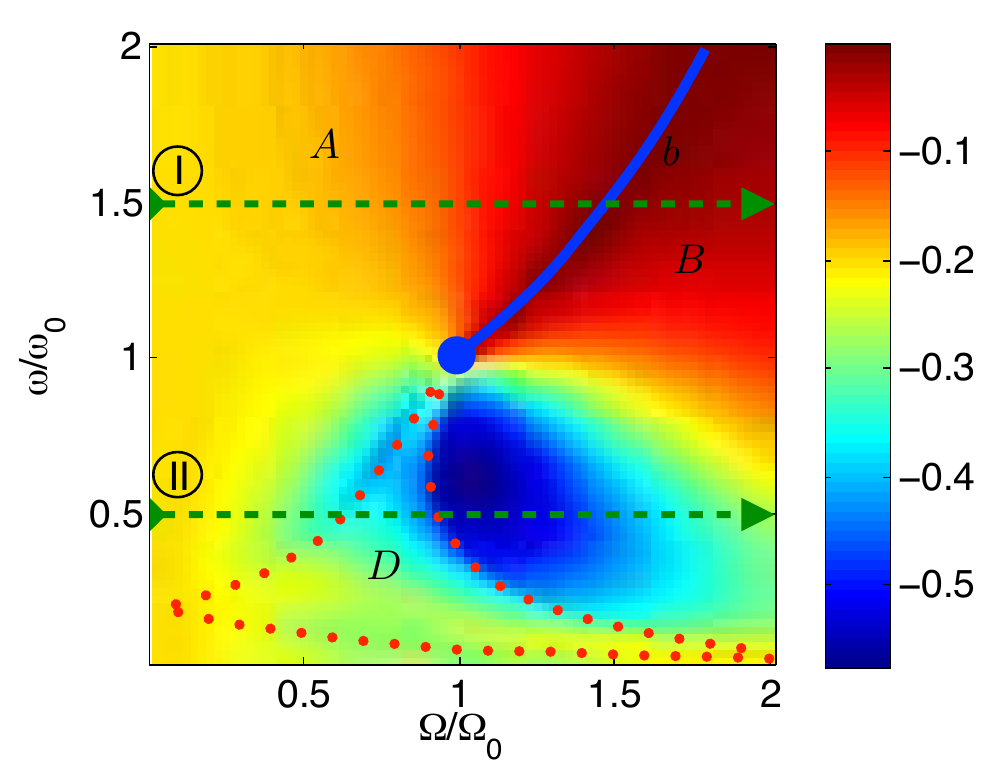}
  \caption{Asymptotic decay rate (ADR, cf. text) for $\gamma=a$ within the perturbative framework. 
  Along $b$ the ADR vanishes non-analytically indicating the stabilizing and destabilizing of the modes of region $A$ and $B$, respectively. 
  $b$ is a first order phase boundary culminating in a second order critical point at $(\omega_0,\Omega_0)$. 
  From here region $D$ opens which is characterized by a non-analyticity in the ADR at a finite value. 
  This indicates a change in the dynamic properties of the system which can not be detected in steady state observables. Within $D$ the system shows an over damped behavior in the vicinity of the steady state.
  }  \label{fig:3}
\end{figure}

The ADR according to the perturbative descriptions based on Gaussian modes is displayed in \figref{fig:3}. It is used to study the transitions involving the Gaussian mode in the thermodynamic limit. 
The ADR vanishes along a line $b$ indicating a phase boundary separating the normal and anomalous spin pumping phase, which is described in Section~\ref{sec:transitions2}. 
Furthermore a non-analyticity of the ADR at a finite value defines region $D$, which characterizes a dynamical phase and is explained in Section~\ref{sec:phases2}.

The dynamical matrix of the first order moments $\Sigma$ provides information
on the stability of the semiclassical solutions, the criticality of the
Liouvillian and the non-analyticities of region $D$. In order to understand
the character of the solutions in the different regions of the phase diagram
we consider next the steady state covariance matrix (CM) of the bosonic system.
For a quadratic evolution like the one of \eqref{eq:effmeq} the steady state covariance matrix contains all information on the state. We deduce the effective temperature and the squeezing of the nuclear spin system, which connects to criticality in the system.

For a one-mode system with vanishing displacements $\E{x}$ and $\E{p}$ [in the
steady state of \eqref{eq:effmeq} this is always the case] the CM is defined as
\begin{align}
\label{eq:covmat1}
\Gamma=
\begin{pmatrix}
2\E{x^2}  &  2\E{xp} - i\\
2\E{px} + i &  2\E{p^2}
\end{pmatrix},
\end{align}
with the usual definitions $x=\frac{1}{\sqrt{2}} (b+b^\dagger)$ and $p=\frac{1}{\sqrt{2}i} (b-b^\dagger)$.
Using \eqref{eq:effmeq} we straightforwardly calculate the steady state covariance matrix $\Gamma_{\textrm{ss}}$ across the phase diagram.
As $\Gamma=\Gamma^T>0$, $\Gamma$ is symplectically diagonalizable, with
 \begin{align}
 \label{eq:covmat}
\Gamma=D O
\begin{pmatrix}
M^2 & 0\\
0  &M^{-2}
\end{pmatrix}
 O^{-1},
\end{align}
where $O$ is orthogonal with $\textrm{det}(O)=1$.
For a single mode, $D\geq 1$ and $M\geq 1$ are real numbers. While $D$ is a measure of the purity of the state [$Tr(\rho^2) = 1/ \sqrt{|\Gamma|}=1/D$], the smallest eigenvalue of $\Gamma$, $\lambda_{\textrm{min}} \equiv D M^{-2}$  determines the amount of squeezing in the system \cite{Kim2002}.  $\lambda_{\textrm{min}}<1 $ indicates squeezing in the bosonic mode. For $M=1$, the covariance matrix \eqref{eq:covmat} describes a thermal state of the bosonic mode and $D$ can be straightforwardly associated to a dimensionless effective temperature 
\begin{align}
T_\textrm{eff} = \textrm{ln}\left[ \frac{2}{\sqrt D -1} +1 \right]^{-1}.
\end{align}
This definition is also meaningful for $M > 1$, since the squeezing operation is entropy-conserving. $T_\textrm{eff}$ is also a measure for the entropy of the spin system, as to leading order it is connected to the bosonic mode via an unitary (i.e., entropy-conserving) transformation. The effective temperature of the different  phases will be discussed below in Sections~\ref{sec:wtl} \&~\ref{sec:phases2} [cf. \figref{fig:temp}].

We stress the point that all properties of the covariance matrix derived within the second order of the perturbative approach are independent of the system size $J$. In particular, the amount of fluctuations (i.e., the purity) in the state does not depend on the particle number. In order to self-consistently justify the perturbative approach, $D$ has to be small with regard to $J$. This implies that in the thermodynamic limit $J\rightarrow\infty$ the perturbative results to second (i.e., leading) order become exact. 
\begin{figure}[h]
  \centering
\includegraphics[width=0.5\textwidth]{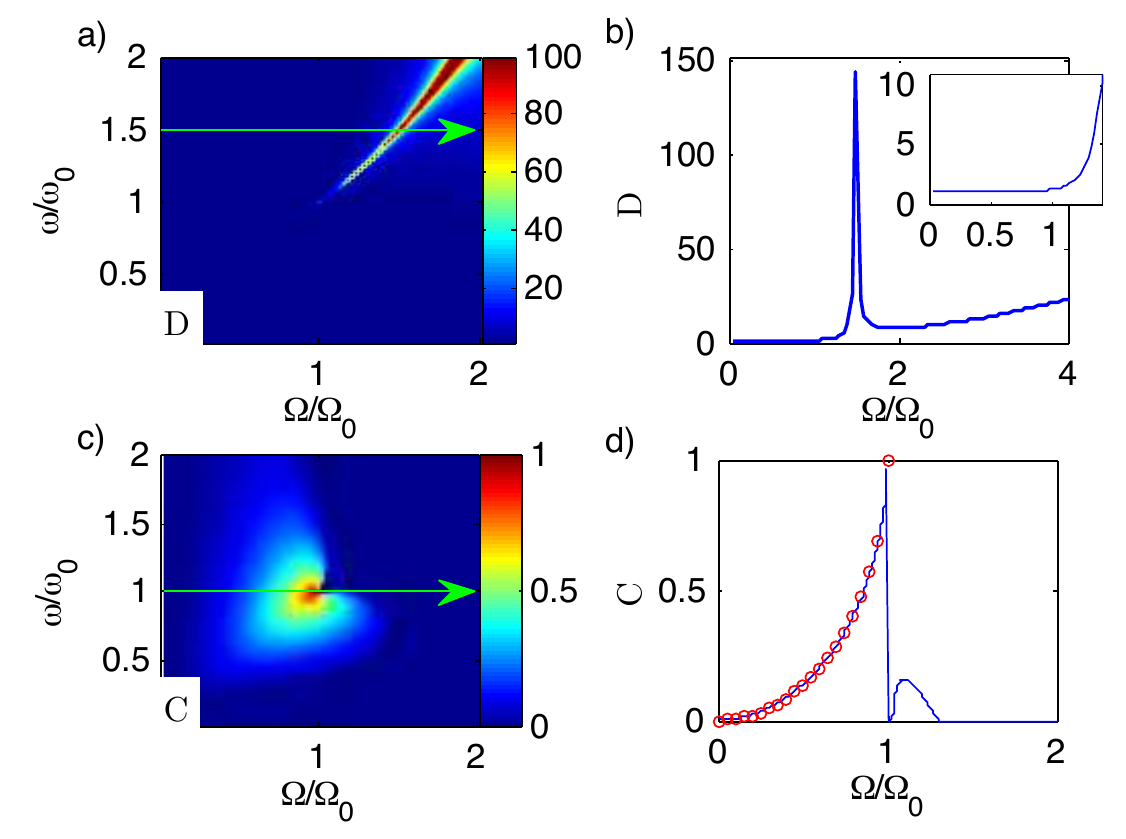}
  \caption{Properties of the steady state covariance matrix $\Gamma_{ss}$ [\eqref{eq:covmat}]. a) The fluctuations $D$ are low in most parts of the phase diagram except for a small wedge around the Gaussian phase boundary. b) Fluctuations $D$ along the line $\omega=1.5 ~\omega_0$ [green line of a)]. The phase boundaries separate a mode with low fluctuations (enlarged in the inset), from a mode with large fluctuations. For large $\Omega$ fluctuations increase, and the system eventually approaches a fully mixed state. c) The squeezing measure $C$ (c.f. text) in the thermodynamic limit. $C$ approaches 1 at $(\omega_0,\Omega_0)$ indicating diverging entanglement in the system. d) $C$ along the line $\omega=\omega_0$ (solid line). The red circles indicate the the squeezing parameter $1-\xi^2_{\hat e_y} = 1- \sqrt{1-(\Omega/\Omega_0)^2}$ (cf. text). 
  }  \label{fig:covmat}
\end{figure}
The inverse purity $D$ is displayed in \figref{fig:covmat}~a).
Except for for a small region around the Gaussian phase boundary $b$ the fluctuations are much smaller than $J=150$, which justifies the validity of the perturbative approach and explains the excellent agreement with the exact diagonalization for this system size.

The squeezing $\lambda_{\textrm{min}}$ in the auxiliary bosonic mode does not necessarily correspond to spin squeezing in the nuclear system. 
In order to deduce the spin squeezing in the nuclear system from the squeezing of the bosonic mode a transformation according to \eqref{eq:J1} and \eqref{eq:Jz1} is necessary. 
In Appendix~\ref{app:exponentialstates} we show that for $|\beta| <1$ \eqref{eq:J1} can be reformulated to connect the spin fluctuations to a  squeezed and rescaled bosonic mode
\begin{align}\label{eq:nanana}
\mathcal J_1^- = \sqrt{2(1-|\beta|^2)} S^\dagger(r) b S(r),
\end{align}
where $S(r)= e^{(r^*b^2 - r b^{\dagger2} )/2}$  is the squeezing operator  and $\textrm{cosh}(r)=\mu = (2k - |\beta|^2)/ [2 \sqrt{2k(1-|\beta|^2)}]$ and $\textrm{sinh}(r) = - \nu =\beta^2/[ 2 \sqrt{2k(1-|\beta|^2)}]$.

Thus squeezing $\lambda_{\textrm{min}}$ of the mode $b$ does in general not imply reduced spin fluctuations in a direction orthogonal to the mean spin polarization since the transformation between spin fluctuations and $b$ involves a squeezing operation itself and  a scaling by a factor $0<\sqrt{2(1-|\beta|^2)}\leq\sqrt 2$.

In general we thus have to apply a more involved squeezing criterion. In
\cite{Korbicz2005} it was shown that for systems of $N$ spin-1/2
particles and for all directions $\vec{n}$ the quantity
\begin{align}
\label{spinsqueezing2}
C_{\vec n} \equiv 1- \frac{2}{J}\E{\Delta I_{\vec n}^2} - \frac{1}{J^2}\E{ I_{\vec n}}^2<1,
\end{align}
signals entanglement if $C_{\vec n}>0$ for some direction $\vec n$. Moreover, $\E{\Delta I_{\vec n}^2}<J/2$
indicates a generalized spin-squeezing of the state
\footnote{In
distinction to the criterion \eqref{eq:spinsqueezing1} the squeezed
component $J_{\vec{n}}$ is not necessarily orthogonal to the mean spin.}.

In the following we will use the quantity $C=\textrm{max}\{0,C_{\vec n}~|\vec n \in \mathbb R^3\}$ to investigate  squeezing and bipartite
entanglement in the nuclear system.
In order to calculate $C_{\vec{n}}$ we reconstruct the approximate
nuclear operators according to \eqref{eq:expansion} and \eqref{eq:Jz}
from the semiclassical displacement $\beta$ and evaluate the expectation
values according to the steady state covariance matrix  \eqref{eq:covmat1}.
Finally we maximize $C_{\vec{n}}$ with regard to all possible directions
$\vec n$ to obtain $C$.
The results are discussed in Section~\ref{sec:transitions2}.
As discussed in more detail in the next Section, the fact that
$C\to1$ as $\Omega\to\Omega_0$ on the line segment $x$ indicates a
 diverging entanglement length in the sense that
$O(1/(1-C))=O(\sqrt J)$-particle entanglement is present \cite{Hyllus2012}.


\subsubsection{A Second Order Phase Transition: The Segment $x$}
\label{sec:wtl}

The segment $x$ at $ \omega=\omega_0$ (\figref{fig:0}) represents a very peculiar region in the phase diagram, where the solution below the critical point can be constructed analytically as seen in Section~\ref{sec:phen}.
The electron and nuclear system decouple, resulting in a zero entropy product steady state. 
A nuclear polarization builds up to cancel the external driving up to the point of maximal Overhauser field ($\Omega_0$).  
At this point squeezing and entanglement in the system diverge, indicating a second order phase transition.  
In the following we exemplarily employ the formalism developed above along this line to obtain further insight about the criticality at $(\omega_0,\Omega_0)$. 
 We calculate the analytical steady state solution as well as the effective master equation governing the nuclear fluctuation dynamics in its vicinity. 
We find that here the spectrum of the Liouvillian becomes continuous (implying a closing gap) and real. At the same time the creation operators of the elementary excitations from the steady state turn hermitian giving rise to diverging spin entanglement. 


  \begin{figure}[h]
  \centering
\includegraphics[width=0.5\textwidth]{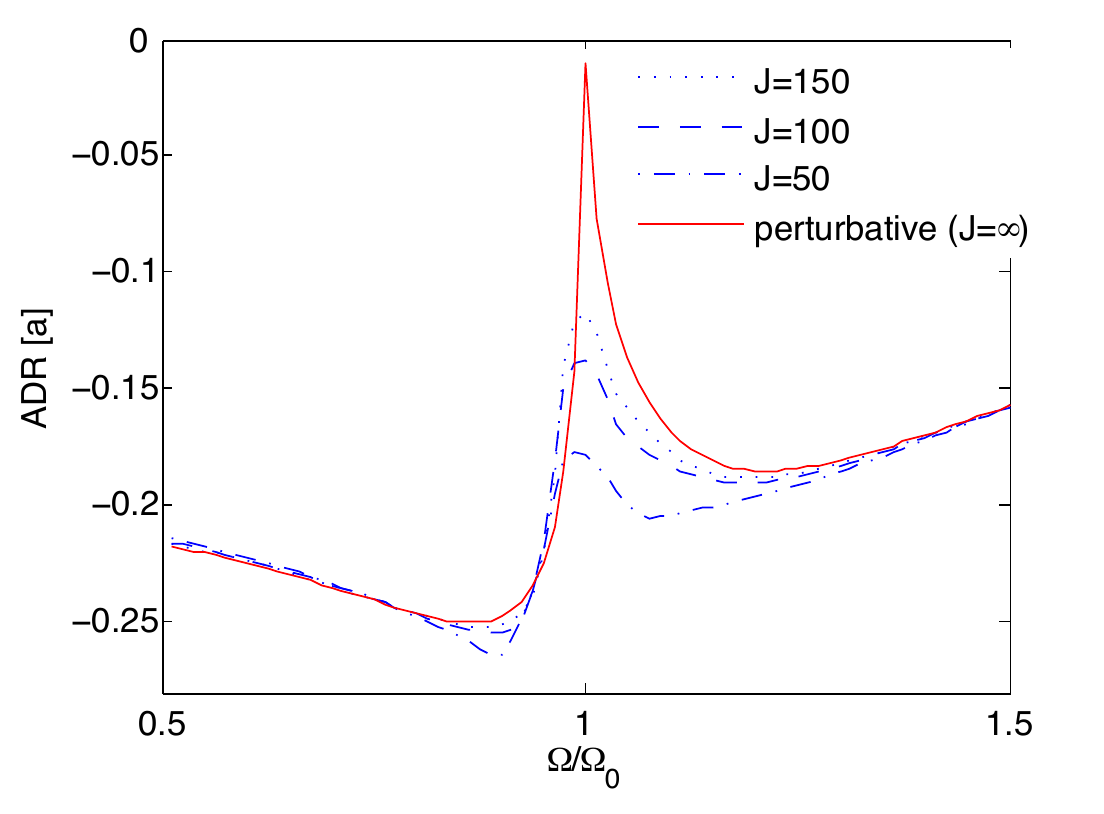}
  \caption{
 The ADR ($\gamma=a$) for $J=50$, 100, 150 (broken lines) in comparison with the perturbatively calculated (solid line, cf. Section~\ref{sec:wtl}) along $\omega=\omega_0$. For finite systems one finds an avoided crossing at $\Omega_0$. The size of the gap reduces with the system size until it closes in the thermodynamic limit (solid line). Below $\Omega_0$ the ADR in the thermodynamic limit is given by \eqref{eq:gammaeff}.
  }  \label{fig:spec_w0}

\end{figure}

The first order stability condition \eqref{eq:beta} is fulfilled, if $\tilde\Omega=0$ [compare \eqref{eq:Z1} and \eqref{eq:Z2}], which yields the possible semiclassical steady state displacements 
\begin{align}\label{eq:disp}
&\sqrt k \beta = -\Omega/\Omega_0 \\\nonumber
\Leftrightarrow&~ \beta_\pm = -\sqrt{1\pm\sqrt{1-(\Omega/\Omega_0)^2}},
\end{align}
 corresponding to a normal ('$-$') and anomalous ('$+$') spin pumping mode, respectively. 

Next, we explicitly calculate the second order corrective dynamics of the nuclear degrees of freedom for the normal mode.
The vanishing of the effective driving $\tilde \Omega=0$ forces the electron in its dark state -- implying $\E{S^+}_{ss}=\E{S^-}_{ss}=\E{S^+S^-}_{ss}=0$ -- and directly yields $B=F=0$ [\eqref{eq:coeff1} and \eqref{eq:coeff11}]. 
The remaining constants can be calculated as described above and
introducing new bosonic operators (for the normal mode $\beta=\beta_-\leq 1$)
\begin{align}
\label{eq:d1}
d=\mu b + \nu b^\dagger,
\end{align}
with
\begin{subequations}
\label{eq:sym}
\begin{align}
\mu = \frac{2k - |\beta|^2}{ 2 \sqrt{2k(1-|\beta|^2)}},\\ \label{eq:2}
\nu  = -\frac{\beta^2}{ 2 \sqrt{2k(1-|\beta|^2)}},
\end{align}
\end{subequations}
one finds the effective evolution of the nuclear fluctuations given as
\begin{align}
\label{eq:meqdw0}
\dot\sigma=& \Gamma_{\mathrm{eff}} \left( d \sigma d^\dagger - \frac{1}{2} \{d^\dagger d ,\sigma \} \right)\\\nonumber
&-i \left[\Theta_{\mathrm{eff}} d^\dagger d,\sigma\right],
\end{align}
with
\begin{align}
\Gamma_{\mathrm{eff}}=2a^2 \mathrm{Re}\left( \frac{1}{\gamma + i 2 a (|\beta|^2 - 1)}\right) (1-|\beta|^2),\\
\Theta_{\mathrm{eff}}=a^2 \mathrm{Im}\left( \frac{1}{\gamma + i 2 a (|\beta|^2 - 1)}\right) (1-|\beta|^2).
\end{align}
$d$ and $d^\dagger$ fulfill boson commutation relations, since \eqref{eq:d1} defines a symplectic transformation ($|\mu|^2-|\nu|^2 =1$). The eigenvalues of the dynamical matrix $\Sigma$ associated to \eqref{eq:meqdw0} are straightforwardly given as $\lambda_{1,2}=-\Gamma_{\textrm{eff}}/2\pm i \Theta_{\textrm{eff}}$. 
The real part -- representing the ADR of the system in thermodynamic limit (compare \figref{fig:spec_w0}) -- is always negative, indicating the stability of the normal spin pumping mode ($\beta_-$).
In an analogous calculation one shows that the semiclassical solution $\beta_+>1$ is not stable to second order since the eigenvalues of $\Sigma$ have a positive real part, i.e. the fluctuations diverge, violating the initial assumptions that the mode $b$ has to be lowly occupied.  

\begin{figure}[h]
  \centering
\includegraphics[width=0.5\textwidth]{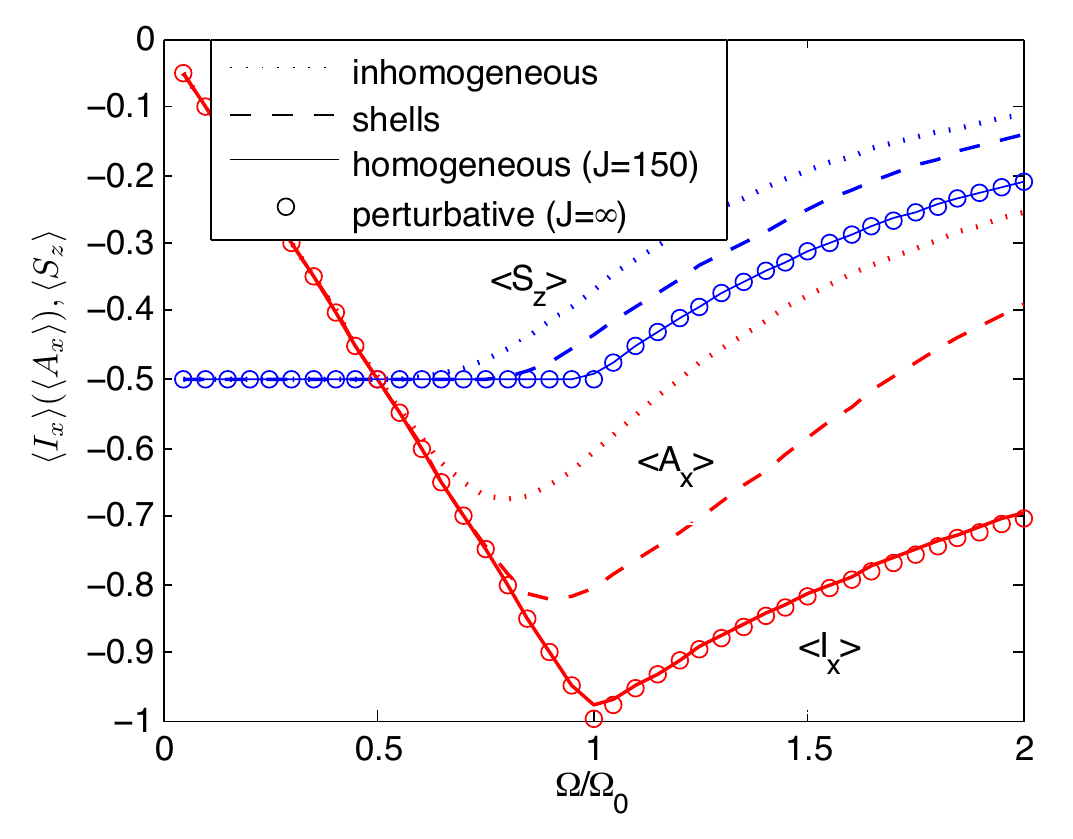}
  \caption{Electron inversion $\E{S_z}$ and the nuclear field in $x$ direction $\E{I_x}$ along $\omega=\omega_0$, in the thermodynamic limit according to the perturbative theory (circles) in comparison with the numeric values from exact diagonalization for a finite system of $J=150$ (solid lines). 
The perturbative theory shows excellent agreement with the numerics.
Further the numerically determined electron inversion and the expectation value of the inhomogeneous nuclear operator $\E{A_x}$ are displayed for a model of two inhomogeneously coupled nuclear shells ($g_1=2g_2$) of size $J_{1,2}=8$ (dashed lines) and for 5 inhomogeneously coupled nuclear spins (dotted lines) are displayed (discussion see Section~\ref{sec:Implementations}).
  }  \label{fig:shells}
\end{figure}

Selected steady state expectation values derived from the stable displacement $\beta_-$ to leading order in $J$ (i.e., in the thermodynamic limit) are displayed in \figref{fig:shells}.
Already for $J=150$ we find excellent agreement between the perturbative and exact mean polarizations. 
The semiclassical nuclear field builds up to exactly cancel the external magnetic field $\Omega$ forcing the electron in its dark state $\ket \downarrow$ along $x$ and thus realizing the model of cooperative resonance fluorescence \cite{HJCarmichael1980} even for weak dissipation $\gamma \leq a$ [compare \eqref{eq:elim}]. This solution is only available if $\Omega \leq \Omega_0$ (defining segment $x$), i.e., up to the point where the nuclear field reaches its maximum. 
At this point the system enters a new phase of anomalous spin pumping (described below) and the steady state properties change abruptly.

Inserting solution $\beta_-$ in the coefficients of Master \eqref{eq:meqdw0} yields
\begin{align}
\label{eq:gammaeff}
\Gamma_{\mathrm{eff}}=2a^2 \mathrm{Re}\left( \frac{1}{\gamma - i 2 a \sqrt{1-(\Omega/\Omega_0)^2}}\right)\sqrt{1-(\Omega/\Omega_0)^2},\\
\Theta_{\mathrm{eff}}=a^2 \mathrm{Im}\left( \frac{1}{\gamma - i 2 a\sqrt{1-(\Omega/\Omega_0)^2}}\right)\sqrt{1-(\Omega/\Omega_0)^2}.
\end{align}
In the close vicinity below the critical point $\Omega_0$ the real part of the gap in the Liouvillian's spectrum closes as
\begin{align}
\label{eq:na1}
\Gamma_{\mathrm{eff}} \approx 2 \frac{a^2} {\gamma} \sqrt{1-(\Omega/\Omega_0)^2},
\end{align}
and the imaginary part as
\begin{align}
\label{eq:na2}
|\Theta_{\mathrm{eff}}| \approx 2 \frac{a^3} {\gamma^2} [1-(\Omega/\Omega_0)^2],
\end{align}
indicating criticality. \figref{fig:spec_w0} displays the ADR along $\omega=\omega_0$ in the thermodynamic limit (which is given on the segment $x$ by \eqref{eq:gammaeff}) and for finite systems. It displays an avoided crossing at $\Omega_0$ with a gap that vanishes in the thermodynamic limit. This closing of the gap coincides with diverging timescales in the system, which renders the model more susceptible to potential perturbing effects, a phenomenon well known in the context of criticality \cite{Bowden1979}.

In contrast to the general form \eqref{eq:effmeq}, \eqref{eq:meqdw0} contains only one Lindblad term and the dynamics drive the system into the vacuum $\ket {0_d}$ of the squeezed mode $d$. As the system approaches the critical value $\Omega =\Omega_0$ (i.e., $\beta_-=-1$) the mode $d$ adopts more and more a $\hat p = \frac{1}{\sqrt{2}i} (b-b^\dagger)$ like character and thus the squeezing of this mode's vacuum increases.
The (in general complicated) transformation between the squeezing of the bosonic mode $b$ and the spin operators (cf. Section~\ref{sec:methods}) can readily be established along $x$, since the  operator $d$ is trivially related to the spin operators [cf. \eqref{eq:J1}]
\begin{align}\nonumber
\mathcal J^-_1&=\frac{1}{2\sqrt{k}} [(2k-|\beta|^2)b- \beta ^2 \bd]\\
&=\sqrt{2(1-|\beta|^2)}(\mu b + \nu b^\dagger)\\\nonumber
&= \sqrt{2(1-|\beta|^2)}d.
\end{align}
The fluctuations in $y$-direction, for example, are consequently given as
\begin{align}
\mathcal J^y_1=\sqrt{(1-|\beta|^2)} \hat p_d,
\end{align}
where $\hat p_d = \frac{1}{\sqrt 2 i} (d-d^\dagger)$.
One readily shows that
\begin{align}
\E{\Delta I_y^2}=J\E{\mathcal J_1^{y2}}=J(1-|\beta|^2)  \E{\hat p_d^2},
\end{align}
up to order $O(1)$ and we used $\E{d}=0$ in the steady state. In the $\hat p$-vacuum $\ket {0_p}$ it is $ \E{\hat p_d^2}=1/2$, such that we  evaluate
\begin{align}
\label{eq:sq}
\xi_{\hat e_y}^2& = 2 \E{\Delta I_{y}^2}/ |\E{\vec I}| \\\nonumber
                           &= 2(1-|\beta|^2)  \E{\hat p_d^2} = \sqrt{1 - \left(\frac{\Omega}{\Omega_0}\right)^2},
\end{align}
where we used $|\E{\vec I}|=J$ and inserted the semiclassical displacement $\beta_-$.

This is the same result we derived in Section~\ref{sec:phen} and Appendix~\ref{app:exponentialstates} by constructing approximate eigenstates of the lowering operator $I^-$ and along $x$ we find that $C\approx 1-\xi_{\hat e_y}^2$, as shown in \figref{fig:covmat}~d). 
Note
that here $\hat e_y$ is orthogonal to the direction of the mean spin
$\E{\vec{I}}$. 
This allows us to deduce that $O(\sqrt J)$
nuclear spins must be entangled close to the critical point, which
establishes a 'diverging entanglement length' in this system. To see
this, we employ a variant of the criterion \eqref{eq:spinsqueezing1} as
discussed in \cite{Pezze2009}. There, it was shown that $\xi_{\hat
e_y}^2<1/k$ sets a lower bound of $N\xi_{\hat e_y}^{-2}$ on the quantum
Fisher information $F_Q$ of the state. In \cite{Hyllus2012} it was shown
that for states containing at most $k$-particle entanglement, $F_Q$ is
upper bounded by $Nk$. Consequently, the values of $\xi_{\hat e_y}^2$ obtained close to the critical point (cf. \eqref{eq:sq}
and Appendix~\ref{app:exponentialstates}) imply that at least
$O(\sqrt J)$-particle entanglement must be present.  
Note that the bosonic description does not allow to describe the range $\xi_{\hat e_y}^2=O(1/J)$, i.e., $k=O(J)$, where the fluctuations become larger than the expansion parameter. 

The nuclear squeezing and entanglement in the system diverges approaching  the critical point, as the Lindblad operator $d$ (defining the steady state $\ket{0_d}$) becomes more and more $\hat p$-like. The fluctuations in $y$-direction tend to zero, while at the same time -- due to the Heisenberg uncertainty relation -- the steady state is in a superposition of an increasing number of $I_z$ eigenstates. Since in a system with infinite range interactions (as the one we are considering) there is no obvious definition of a coherence length,  the range of the involved $I_z$ eigenstates can be considered as an analogous concept.

At the critical value $\Omega =\Omega_0$ the symplectic transformation Eqs.~(\ref{eq:d1}) becomes ill defined ($d$ becomes a $\hat p$-like operator) while both the dissipation rate and the mode energy tend to zero. While the coefficients in Eqs.~(\ref{eq:sym}) diverge, the total master equation is well defined [due to the factors $ (1-|\beta|^2)$ in $\Gamma_{\mathrm{eff}}$] and straightforwardly can be written as
\begin{align}
\label{eq:meqcrit}
\dot\sigma=&\frac{a^2}{2\gamma} \left( \hat p \sigma \hat p - \frac{1}{2} \{\hat p^2 ,\sigma \} \right).
\end{align}
The Liouville operator's spectrum is real and continuous with hermitian creation operators of the elementary excitations. 

We stress the point that along segment $x$ in the phase diagram highly dissipative dynamics drive the system in a pure and separable steady state with zero effective temperature $T_\textrm{eff}=0$ [cf. \figref{fig:temp}~b)]. At the critical point $\Omega_0$ the steady state changes its nature abruptly as the system enters a high temperature phase.

Furthermore we remark that this steady state has no relation to the system's ground state. This is in contrast to the extensively studied Dicke phase transition \cite{Baumann2010a,Nagy2010,Lambert04} where the steady state is in close relation to the Hamiltonian's ground state (in fact in the normal phase it is identical). In the present model dissipation drives the system to a highly excited state of the Hamiltonian and the observed critical phenomena are disconnected from the Hamiltonian's low excitation spectrum.

We have seen that at the critical point ($\omega_0,\Omega_0$) the gap of the Liouville operator's spectrum (in both real and imaginary part) closes in the thermodynamic limit [\eqref{eq:na1} and \eqref{eq:na2}]. 
Approaching the critical point the steady state fluctuations become more and more squeezed due to the increasing $\hat p$-like character of the mode $d$. 
The spin squeezing close to the critical point [\eqref{eq:sq}] can be interpreted as a diverging coherence length in a system with infinite range interactions (the electron mediates interactions between remote spins). 
These are clear indications for a second order phase transition, which will be formalized in Section~\ref{sec:transitions2}.


\subsubsection{Phases}
\label{sec:phases2}

In the present Section we study the different phases of the system, which involve the RSTSS solution ($A$, $B$ and $D$) using the analytic tools developed above. 
By construction, the RSTSS solution describes steady states where the electron and nuclear state factorize to leading order in the system size and the nuclear system is found in a fully polarized and rotated state with Gaussian fluctuations, which are fully characterized by their effective temperature and squeezing. 
\figref{fig:1} displays different steady state observables of the Gaussian solution determined via the formalism described above to leading order in the thermodynamic limit.

  \begin{figure}[h]
  \centering
\includegraphics[width=0.5\textwidth]{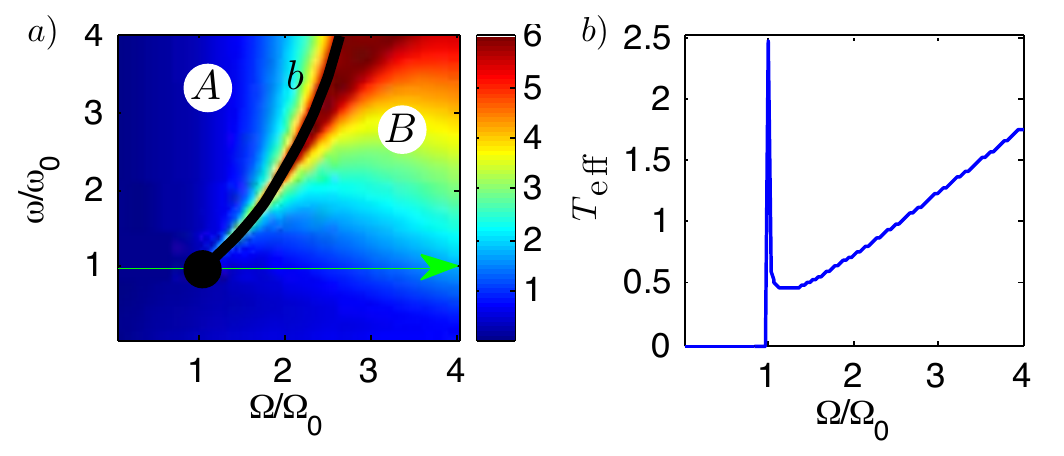}
  \caption{Effective temperatur $T_{\textrm{eff}}$ of the gaussian mode. Temperatures $T_\textrm{eff}>6$ are cut off, as the temperature diverges along the phase boundary $b$.
a) The first order phase boundary $b$ separates the low temperature phase $A$ from the high temperature phase $B$. b) $T_\textrm{eff}$ along $\omega=\omega_0$: On segment $x$ the system is in a zero entropy state ($T_{\textrm{eff}}$=0). Above the second order critical point $\Omega>\Omega_0$ the system enters a high temperature phase. Here the temperature rises with increasing driving strength.
  }  \label{fig:temp}

\end{figure}

In phase $A$ the system is characterized by normal spin pumping behavior. Only the semiclassical displacement $\beta_-$ (normal mode) leads to a dynamical matrix $\Sigma$ that has negative real parts of its eigenvalues, while for $\beta_+$ the eigenvalues have positive real parts, indicating the instability of that mode in second order. 
The nuclear system in the normal mode settles in a state highly polarized in -$z$ direction following the direction of the electron spin pumping [\figref{fig:1}(a)]. Meanwhile, increasing the external driving $\Omega$ and approaching the phase boundary $b$, a nuclear field in $x$ direction builds up, but only along $x$ it can fully cancel the external driving [\figref{fig:1}(c)].
Therefore, in general the electron spin aligns more and more with the external field [\figref{fig:1}(b,d)]. Furthermore, the effective temperature (and thus the entropy) of the phase is low, as displayed in \figref{fig:temp}~a).

In region $B$ in contrast, $\beta_+$ is the only stable solution, defining the phase of anomalous spin pumping behavior. The nuclear system now shows strong population inversion, i.e., the nuclear polarization is in direction opposite to the external pumping ($z$). In the same way the electron now aligns in opposite direction to the external driving field ($x$). Also, in contrast to phase $A$, the RSTSS now is in a high-temperature state. 
For larger electron driving the temperature increases until eventually the Gaussian description breaks down (as $D\propto J$) and for $\Omega \rightarrow \infty$ the system is found in a completely mixed state [compare \figref{fig:covmat}~b)].

In the upper half of the phase diagram ($\omega > \omega_0$) phase $A$ changes abruptly into phase $B$ at the boundary $b$ and certain steady state spin observables [$\E{I_z}$, $\E{S_x}$ [\figref{fig:1}~a) \& b)] and $\E{I_y}$ (not displayed)] show distinct features of a first order phase transition, changing sign as the normal (anomalous) mode destabilizes (stabilizes). 
This transition is discussed in greater detail in the following Section~\ref{sec:transitions2}.
Following this boundary towards the critical point ($\omega_0,\Omega_0$) the two phases become progressively more similar. Below the critical point ($\omega < \omega_0$) there is no clear distinction between the normal and anomalous spin pumping mode anymore, a phenomenon known from thermodynamics as \textit{supercriticality}. Phase $A$ transforms continuously to phase $B$ in this region. Close to the critical point, supercritical media typically respond very sensitive to the external control parameters of the phase diagram (e.g., temperature or pressure) \cite{Clifford:1999uv}. In our system we observe that small changes in the parameter $\omega$ leads to large changes in electron spin observables.



Next, we consider the third region associated with the RSTSS solution, region $D$. We will find that this region differs from the previous ones by the fact that it cannot be detected in the system's steady state but rather in dynamical observables.

The eigenvalues of the dynamic matrix $\Sigma$ can be calculated as $\lambda_{1,2}=-(R_a-R_b) \pm 2 \sqrt{4 |\xi|^2 - \chi^2 }$ and provide information on the approximate low excitation spectrum of the Liouvillian. We can distinguish two cases for the low excitation spectrum, which differ only in the Hamiltonian properties of \eqref{eq:effmeq} (fully determined by $\chi$ and $\xi$ [\eqref{eq:chi} \& \eqref{eq:xi}]).
In the first case the quadratic bosonic Hamiltonian can be symplectically transformed to be diagonal in a Fock basis (i.e., of the form $\propto \tilde b^\dagger \tilde b$). This is the case if $\chi^2 > 4 |\xi|^2 $. As a consequence the two eigenvalues of $\Sigma$ have an identical real part
and imaginary parts $\pm  2 \sqrt{ \chi^2-4 |\xi|^2  }$. 
In the second case the Hamiltonian transforms symplectically into a squeezing Hamiltonian $\propto  (\tilde b^{\dagger2} + \tilde b^2$). Here one finds $\chi^2 < 4 |\xi|^2 $, such that the eigenvalues become real and symmetrically distributed around $-(R_a-R_b)$.
In region $D$ in \figref{fig:0} we find the effective Hamiltonian for the nuclear fluctuations to be symplectically equivalent to a squeezing Hamiltonian. 


\begin{figure}[h]
  \centering
\includegraphics[width=0.5\textwidth]{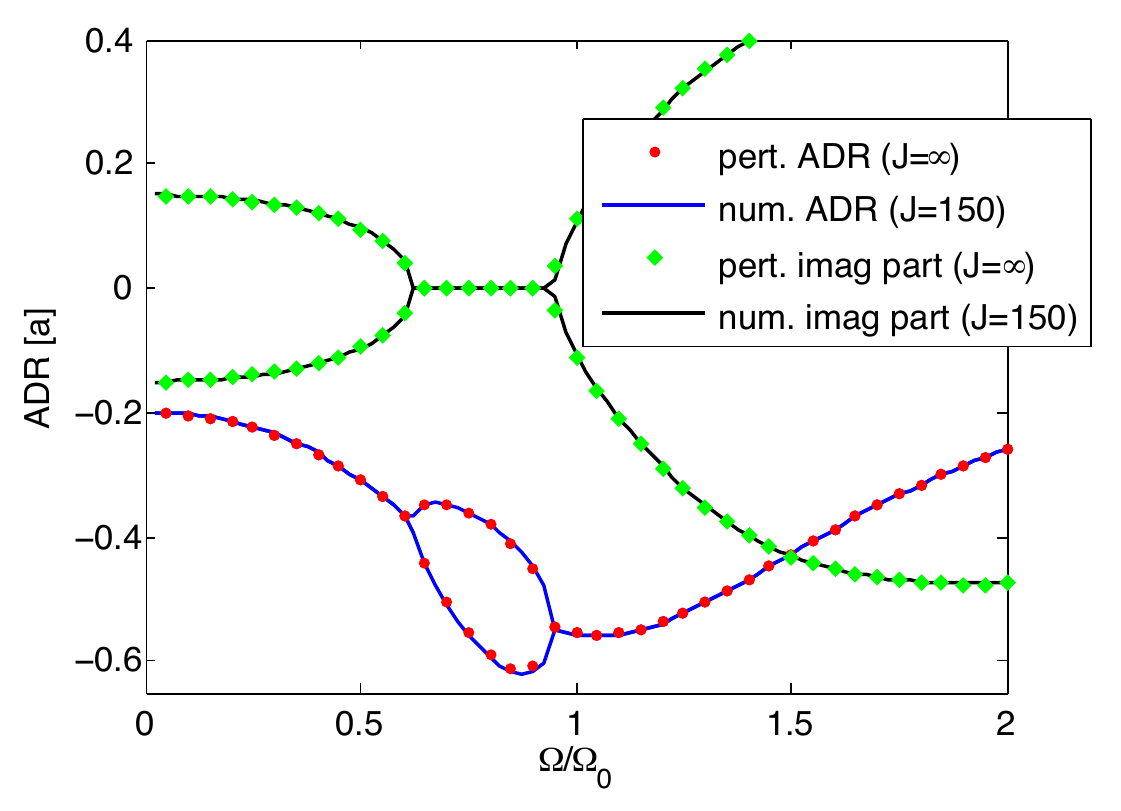}
  \caption{ The ADR and the imaginary part of the respective eigenvalue ($\gamma=a$) for $J=150$ (solid
    lines) in comparison with the perturbatively calculated value (dots) along \textcircled{\tiny{II}} of \figref{fig:3}. In the region where the coherent part of \eqref{eq:effmeq} is a squeezing Hamiltonian, the ADR (i.e. real part of the lowest Liouvillian eigenvalue pair) splits. At the same time the imaginary part of the lowest eigenvalue pair vanishes (black dashed lines), indicating that the system is overdamped.
  }  \label{fig:ADR05}
\end{figure}

\figref{fig:ADR05} shows the ADR exemplarily along the line $\omega=0.5~\omega_0$ (~\kreis{\footnotesize{II}} in \figref{fig:0}) calculated according to the perturbative theory and via exact diagonalization, respectively. 
The perturbative theory approximates accurately the low excitation spectrum of the Liouvillian. 
We find that in region $D$ the ADR splits up, when the coherent part of \eqref{eq:effmeq} changes to a squeezing Hamiltonian.
As mentioned above this non-analyticity occurs at a non-zero value of the ADR and thus does not leave signatures in the steady state behavior. 
The steady state transforms smoothly along  ~$\kreis{\footnotesize{II}}$.
However the nature of dynamical observables change within region $D$ as the system displays anomalous behavior approaching the steady state. 
The splitting of the ADR coincides with the vanishing of the imaginary part of the lowest non-zero Liouvillian eigenvalues. 
Thus the system is overdamped in $D$. Perturbing the system from its steady state will not lead to a damped oscillatory behavior, but to an exponential, oscillation-free return to the steady state.

The blue area in vicinity to region $D$ in \figref{fig:3} does not represents a new phase but is another interesting feature of the system. Here, the ADR exceeds the value at $\Omega=0$ by a factor $\sim 3$. For $\Omega=0$ the model describes the standard spin pumping setting. 
Large gaps in the low excitation spectrum indicate the possibility to improve the effective spin pumping rate (remember that also in this region the steady state is fully polarized, however not in $-z$ direction as it is the case for the normal spin pumping configuration $\Omega=0$). Indeed simulations show that starting from a fully mixed state, the system reaches the steady state faster than in the standard setting ($\Omega=0$). This feature becomes more distinct in systems, where the electron pumping rate $\gamma$ is limited. For $\gamma =0.1 a$ the time to reach the fully polarized steady state from a fully mixed state is shortened by a factor $\sim 6$.

\subsubsection{Transitions}
\label{sec:transitions2}

In this Section we consider the transitions involving the RSTSS solution in
greater detail providing a classification in analogy to quantum phase transitions in closed systems (compare Section~\ref{sec:frame}). 

\begin{figure}[h]
  \centering
\includegraphics[width=0.5\textwidth]{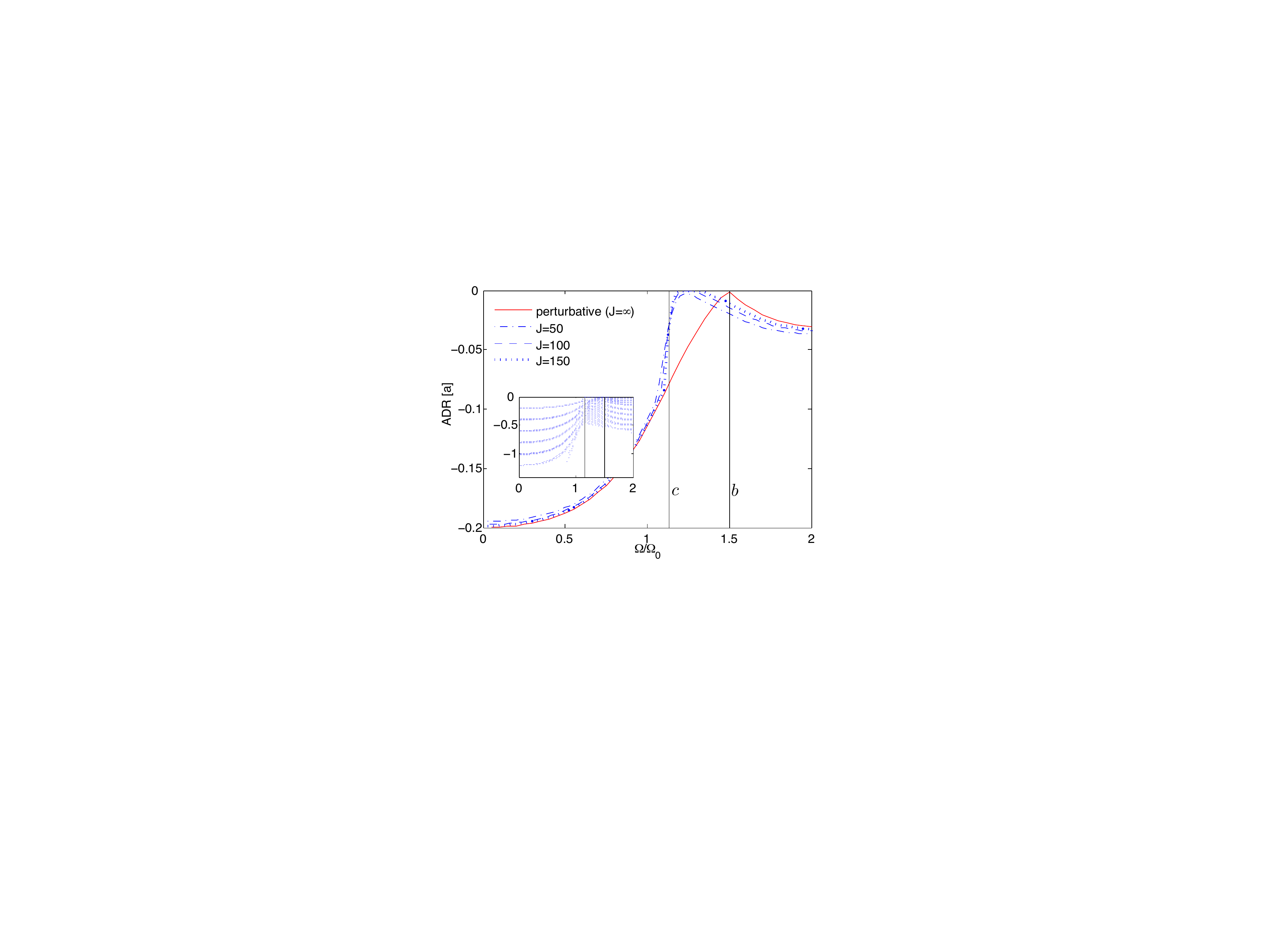}
  \caption{
 The ADR ($\gamma=a$) for $J=50$, 100, 150 (broken lines) in comparison with the perturbatively calculated (solid line) along \textcircled{\tiny{l}}  of \figref{fig:3}. The vertical black lines indicate the asymptotic boundaries of the region of bistability. In the whole region the ADR tends to zero in the thermodynamic limit due to the appearance of a non-gaussian stable mode.  
 \textit{Inset:} The next higher excitations in the spectrum for $J=150$ display equidistant splittings in regions far from the region of bistability. This is an indication for the bosonic character of the steady state, which is exploited in the perturbative approach.
  }  \label{fig:4}
  \end{figure}

As seen in the previous Section, certain steady state observables show clear signatures of a first order phase transition at $b$ (\figref{fig:1}). 
In order to understand this sharp transition we consider the ADR exemplarily along path ~$\kreis{\footnotesize{I}}$ in \figref{fig:4}. 
The broken lines represent numeric results of exact diagonalization of the Liouvillian for $J=50$, $100$ and $150$, while the solid line indicates the result of the perturbative approach. As described in Section~\ref{sec:methods} we implicitly choose the semiclassical displacement $\beta_-$ (for $\Omega< 1.5 \Omega_0$) or $\beta_+$ (for $\Omega> 1.5 \Omega_0$) for which the ADR is negative, indicating a stable solution. 
For increasing system size the ADR is increasingly well approximated by the perturbative solution. 

We stress the point that the red line represents the first Gaussian excitation energy only. However, within the region of bistability (indicated by two vertical bars and discussed below in Section~\ref{sec:RoB}), a non-Gaussian mode (discussed below) is responsible for additional excitations in the exact spectrum. 
The Gaussian mode eigenvalue (red line) in this region is reproduced approximately by higher excitations of the exact spectrum (not displayed) . The perturbative theory is still correct within the region of bistability but, as expected, it misses all non-Gaussian eigenstates of the exact Liouvillian.

At the boundary $b$ ( $\Omega \approx 1.5~\Omega_0$ ) the gap in the real part of the spectrum of the Liouvillian closes non-analytically, indicating critical behavior. This observation is supported by the effective temperature (and thus the fluctuations in the system), which is increased in the vicinity of the boundary $b$, and diverges at the boundary [\figref{fig:temp}~a) \& \figref{fig:covmat}~a)]. 
The vanishing of the ADR at $b$ (i.e., the vanishing due to the RSTSS solution) can be observed at finite $J$ (dashed lines in \figref{fig:4}) and is not a feature appearing in the thermodynamic limit only. The position of this closing of the gap -- which in the thermodynamic limit (solid line) is found at $\Omega \approx 1.5~\Omega_0$ -- is shifted for finite system sizes to lower drivings $\Omega$. 

The origin of this closing of the Liouvillian gap becomes more transparent if we take the mode energy  of the respective metastable solution into account.

\begin{figure}[h]
  \centering
\includegraphics[width=0.5\textwidth]{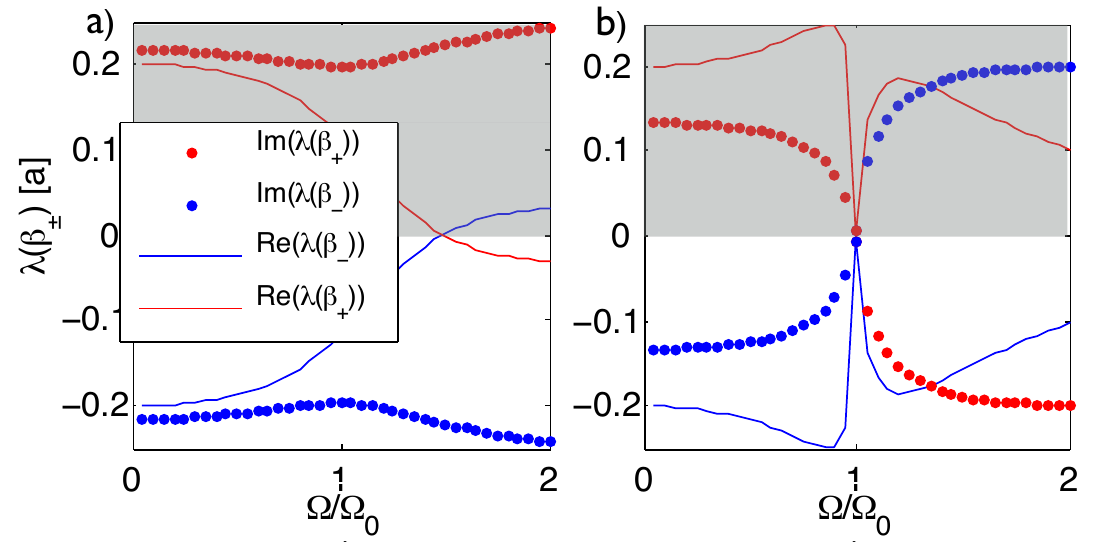}
  \caption{Complex energy of the two modes corresponding to the semiclassical solutions $\beta_\pm$ for $\gamma=a$. The solid line in the non-shaded area represents the ADR of \figref{fig:4} and \figref{fig:spec_w0}, respectively. a) Along \textcircled{\tiny{l}} ($\omega= 1.5 ~\omega_0$). The eigenvalues miss each other in the complex plane. The real parts cross directly. b) $\omega=  \omega_0$. The eigenvalues degenerate asymptotically (in both real and imaginary parts) at the critical point. This closing of the gap originates from an avoided crossing in finite systems with the relevant gap vanishing in the thermodynamic limit (see also \figref{fig:spec_w0})  }  \label{fig:pertspect}
\end{figure}

In \figref{fig:pertspect}~a) the complex energy of both the stable and unstable mode are displayed (i.e., the first eigenvalue of the matrix $\Sigma$ [\eqref{eq:b}]). 
The normal spin pumping mode ($\beta_-$; blue lines) is stable [$\textrm{Re}\lambda(\beta_-)<0$]
up to the critical point where it destabilizes and the anomalous mode appears ($\beta_+$; red lines).
At the critical point the two solutions are macroscopically different $\beta_-\neq\beta_+ $ and their energy [Im($\lambda{\beta_\pm}$)] is distinct across the transition [dotted lines in \figref{fig:pertspect}~a)]. 
Although the projection of the eigenvalues on the real axis vanishes at the critical point for both modes (indicating the stabilizing / destabilizing of the modes) the eigenvalues pass each other in the complex plane at large distance. 
There is no degeneracy in the spectrum of the Liouvillian at the critical point and consequently there can be no mixing of the two modes; the real parts of the eigenvalues cross \textit{directly} without influencing each other. 
Except for the change in stability the modes do not change their character approaching the phase boundary and no diverging correlations (indicated by the squeezing parameter $C$) can be observed.
Together with the discontinuous change in system observables such as mean polarizations we classify this Gaussian transition as of first order.

Second, we consider the transition along $\omega=
\omega_0$ (including the line segment $x$).  
In contrast to the situation before we find that the semiclassical displacements $\beta_+$ and $\beta_-$ merge approaching the critical point such that the two modes become asymptotically identical at $\Omega_0$ [\eqref{eq:disp}].
Approaching the critical point, the eigenvalues of the two modes tend to zero (both the real and imaginary parts), causing the gap of the Liouvillian's spectrum to close [\figref{fig:pertspect}~b), \eqref{eq:na1}, \eqref{eq:na2}]. 
As we have seen in Section~\ref{sec:wtl} at ($\omega_0,\Omega_0$) the spectrum becomes real and continuous signaling criticality.
The perturbative treatment intrinsically is a description in the thermodynamic limit. 
If we consider the exact spectrum we indeed find an avoided crossing due to the mode mixing at the critical point with a gap that is closing for $J\rightarrow \infty$ (cf. \figref{fig:spec_w0}). 
As we discussed in Section~\ref{sec:wtl} the elementary excitations become $\hat p$-like, causing a diverging coherence length in the system [indicated by the diverging squeezing parameter $C$ in \figref{fig:covmat}(c,d)]. 
Together with the continuous but non-analytical change of the mean polarizations these properties classify the point ($\Omega_0,\omega_0$) as a second order transition.

\section{Region of bistability:  Non-Gaussian Solution}
\label{sec:RoB}

As noted in Section~\ref{sec:phen} along the Gaussian boundary $b$ extends a region of bistability [$C$ in (\figref{fig:0})] -- culminating in the critical point $(\Omega_0,\omega_0)$ -- in which a second stable solution appears. 
Within the perturbative framework from Section~\ref{sec:PS} this highly non-Gaussian solution could not be detected because it features large fluctuations of the order of the system size $J$. 
In the following we use numerical techniques to construct and study this mode for finite systems.
In the thermodynamic limit the ADR tends to zero within $C$, such that there exists a two dimensional subspace of steady states. 
Here we find two independent, physical solutions within the two dimensional kernel of the Liouvillian, one of which will turn out to be the Gaussian normal spin pumping mode described in Section~\ref{sec:PS}.
We analyze the nature and properties of the other, non-Gaussian solution, exemplarily along the line $\omega= 1.5 ~\omega_0$ (~\kreis{\footnotesize{I}} in \figref{fig:0}).


\figref{fig:4} displays the ADR for different particle numbers. 
Within the indicated region of bistability (the black vertical lines represent the boundaries $c$ and $b$, respectively) the ADR tends to zero with increasing particle number.
Already for $J=150$ one finds a small region, where the ADR is small enough (of the order of $10^{-6} a$) that one can construct two linearly independent (quasi) steady state solutions. 
Although we find the eigenmatrix $\rho_1$ associated with the ADR to be non-positive and traceless (the latter being a consequence of $\mathcal L$ being the generator of a trace preserving map) we can linearly combine it with the true steady state $\rho_0$ to obtain two linear independent, positive solutions with trace one, $\rho_{\textrm{lo}}$ (corresponding to the normal spin pumping mode) and $\rho_{\textrm{up}}$. 
These solutions span the two dimensional space of steady states in that region. 
\begin{figure}[h]
  \centering
\includegraphics[width=0.5\textwidth]{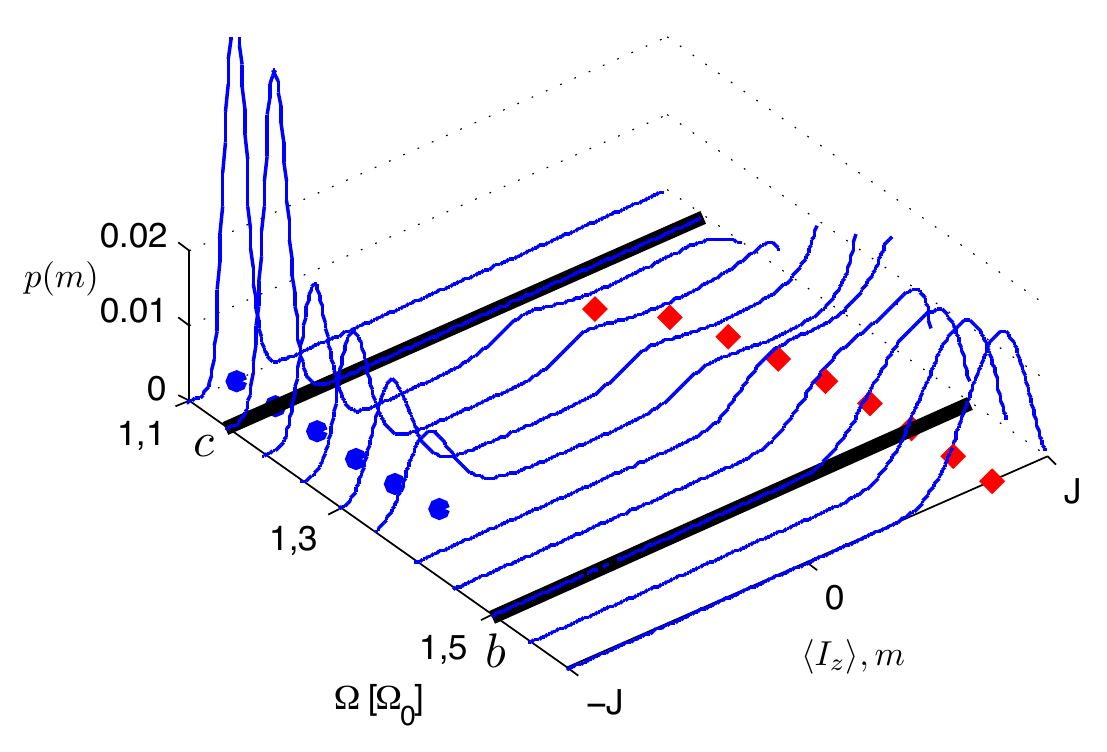}
  \caption{ Diagonal elements $p(m) = \bra m \rho\ket m$ of the nuclear density matrix in $z$-basis ($I_z\ket m =m \ket m$) across the region of bistability for $J=150$, $\gamma=a$. In the bistable regions two quasi stable modes -- the Gaussian normal spin pumping mode (lower branch; $\rho_{\textrm{lo}}$) and a non-Gaussian (upper branch; $\rho_{\textrm{up}}$) -- coexist. The blue dots (red diamonds) in the plane indicate the average polarization in $z$-direction $\E{I_z}$ for the lower (upper) solution.
  }  \label{fig:5}
\end{figure}

\figref{fig:5} illustrates the solutions $\rho_{\textrm{lo}}$ and $\rho_{\textrm{up}}$ around the bistable region in an equally weighted mixture. 
The density matrices are represented by their diagonal elements in the $I_z$ basis. 
In the plane the blue dots (red diamonds) represent the polarization in $z$-direction $\E{I_z}$ of the lower (upper) solution $\rho_{\textrm{lo}}$ ($\rho_{\textrm{up}}$). 
Coming from below the critical region ($\Omega < 1.15~ \Omega_0$) the nuclear system is found in the Gaussian normal spin pumping mode, fully polarized, slightly rotated away from the $-z$ direction and with fluctuations of the order of $\sqrt{J}$. 
This Gaussian solution persists within the critical region where it becomes noisier until eventually -- approaching the right boundary $b$ at $\Omega=1.5~\Omega_0$ -- it destabilizes. 
In the thermodynamic limit the lower solution is stable up to the right boundary, where a first order transition occurs and the anomalous spin pumping mode appears.
Approaching boundary $b$ from above ($\Omega >1.5~ \Omega_0$) this mode transforms into a non-Gaussian solution, which -- in contrast to the coexisting normal mode -- features fluctuations of the order of $J$, is not fully polarized and shows large electron-nuclear and nuclear-nuclear connected correlations. 
Approaching the left boundary $c$ at $\Omega=1.15~\Omega_0$ this mode destabilizes eventually as the ADR becomes finite again and the normal mode is the only stable solution in the system. 

The bistable behavior of the system in region $C$ bears close resemblance to the phenomenon of optical bistability for saturable absorbers \cite{Drummond1980}, where connections to phase transitions have been established \cite{Bowden1979}.
In this region the system displays strong hysteretic behavior. Recent experiments in quantum dots, realizing a setting close to our model system display distinct signatures of hysteresis upon application of an external driving field on the electronic spin \cite{Krebs2010,RFTS07}. Our results suggest the observed optical bistability in central spin systems as a possible pathway to understand these experimental results, which will be a subject of further studies.

\section{Implementations and extensions of the Model}
\label{sec:Implementations}
In the present section we discuss potential physical realizations of the Master \eqref{eq:meq} and 
address certain aspects of an extension of the model for inhomogeneous hyperfine couplings. 

As mentioned above the model we study is a generic central spin model with various potential physical implementations. The most prominent ones represent singly charged semiconductor quantum dots, where the electron spin couples to the nuclear spins of the host material \cite{SKL03,Urbaszek2012}, and diamond nitrogen vacancy centers coupled to either nuclear ($ ^{13}C$ spins of the host material) or electron (e.g., nearby nitrogen impurities) spin ensembles \cite{Jelezko:2006,Togan:2011}.  Recently diamond nano-crystals containing single NV centers coated with organic molecule spin labels, which are dipole coupled to the NV center spin have been manufactured \cite{Chisholm:2011uj}. 

NV centers represent a natural realization of the Master \eqref{eq:meq}. 
Their ground state consists of three spin sublevels (of spin projection quantum number $m=0,\pm1$) featuring a zero field splitting due to anisotropic crystal fields of \SI{2.88}{GHz} \cite{Jelezko:2006}. 
In a static magnetic field this zero field splitting can be compensated for and one of the transitions (e.g., $m=0 \leftrightarrow1$) is brought into near hyperfine resonance with the ancilla spin system, defining an effective two-level system. 
Since the $m=0$ level does not carry a magnetic moment, the hyperfine interaction of the effective two level system and the ancilla system takes the anisotropic form of \eqref{eq:hyp}. Potential counterrotating terms of the dipole-dipole interaction are neglected in the static magnetic field in a rotating wave approximation. 
Optical pumping of the electron spin in the $m=0$ spin state and resonant driving (either by optical Raman transitions or radio frequency fields) realizes Master \eqref{eq:meq} \cite{Tamarat:2008ko}.

In general the hyperfine interaction in such a setting will not be homogeneous and the truncation to a symmetric subspace of total spin $J$ is not justified. 
In the following we consider an extension of the model taking into account the inhomogeneous nature of the hyperfine coupling in a shell model. 
Along $x$ we show that up to the critical point steady states can be constructed analytically as factorized product states involving nuclear eigenstates of the (inhomogeneous) lowering operator. In analogy to the homogeneous case, such solutions cease to exist after the critical point at which we find diverging nuclear squeezing. These results are supported by numerical simulations that confirm the analytical considerations and provide further indications that other features of the phase diagram aside from the second order transition can be found in the inhomogeneous model. 

In order to take into account inhomogeneities in the hyperfine coupling, we replace the homogeneous spin operators of \eqref{eq:hyp} with inhomogeneous operators $I_\alpha\rightarrow A_\alpha$ ($\alpha=x,y,z$). 
We approximate the actual distribution of coupling strengths by $n$ shells of spins with identical coupling 
\begin{align}
A_\alpha=\sum_{i=1}^n g_i A^{(i)}_\alpha,
\end{align}
where $A^{(i)}_\alpha$ represent homogeneous spin operators within the $i$th shell. Each homogeneous shell is assumed to be in a symmetric subspace $J_i$.

In analogy to the homogeneous case we can construct approximate eigenstates of the lowering operator $A^-\ket \alpha = \alpha \ket \alpha$. To this end we perform a Holstein-Primakoff transformation on the homogeneous spin operators within each shell and displace the respective bosonic mode $b_i$ by $\beta_i$ and expand the resulting operators in orders of $1/\sqrt{J_i}$. 
As we demonstrate in Appendix~\ref{app:inhomo} the choice of a particular displacement $\beta_i$ uniquely defines the squeezing of the respective mode $b_i$ if we demand that the corresponding state is an  $A^-$ eigenstate to second order in the expansion parameters, i.e., of order $\mathcal O(\sum_i 1/J_i)$. The corresponding eigenvalue is then given as $\alpha =\sum_{i=1}^n g_i \sqrt{k_i} \beta_i$ ($k_i=2-|\beta_i|^2$). As discussed in Section~\ref{sec:phen}, $\ket \psi = \ket \downarrow \otimes \ket \alpha$ is a steady state of the evolution to second order, if $\alpha =\sum_i g_i \sqrt{k_i} \beta_i = -J\Omega/\Omega_0$. In contrast to the homogeneous case ($n=1$) the latter condition does not determine the steady state uniquely. Several sets of displacements within the different shells can fulfill the steady state condition. However, all these microscopic realizations lead to the same macroscopic behavior of the system such as the locking of the electron inversion $\E{S_z}=0$. Furthermore, at the critical point, the solution is unique again ($\beta_i=1$ for all shells) and the considerations on entanglement of Appendix~\ref{app:exponentialstates} can be straightforwardly generalized to the inhomogeneous case with the result that also here at the critical point the entanglement in the system diverges indicating a second order phase transition. Obviously, above the critical point no such solution can be constructed and the system observables change non-analytically.  

\figref{fig:shells} shows numerical results which confirm the above considerations. We find numerically the exact steady state solution for a model of two inhomogeneously coupled shells ($g_1=2g_2$) of size $J_{1,2}=8$ (broken lines), as well as for a system of 5 nuclear spins with coupling strengths ($\{g_i\}_{i=1\dots5}=\{0.67,0.79,0.94,1.15,1.4\}$, dotted lines).  For low driving strengths $\Omega$ we find the Overhauser field building up linearly, as expected. The emergence of the thermodynamic phase transitions can be anticipated already for these low particle numbers. 

These analytical and numerical arguments for the emergence of a second
order phase transition in the inhomogeneous case, suggest the
possibility to find other features of the homogeneous phase diagram also in
inhomogeneous systems, such as NV centers in diamond.

Another attractive realization of a central spin system is provided by
singly charged semiconductor quantum dots: up to several $10^4$ nuclear
spins are coupled to a central spin-1/2 electron, driving and spin
pumping of the electronic state have been demonstrated experimentally with high efficiency \cite{Kroner:2008kf,Atature:2006ha}.
In this setting, however, the inhomogeneity of the hyperfine coupling
and the absence of an $m=0$ central spin state lead to a situation in
which the effective nuclear Zeeman term $H_I$ in Eq.~(\ref{eq:meq})
becomes inhomogeneous (it is composed of Knight field, nuclear Zeeman
energy, and the (homogeneous) detuning) and does not vanish for any choice of parameters. Therefore the
above argument for a persistence of the second order phase transition
does not apply. However, critical phenomena similar to the ones
described above were observed in optically driven quantum dots
\cite{Krebs2010}. The adaptation of our model this and other more
general settings is subject to future studies.

\section{Conclusions}
\label{sec:Conclusions}

In analogy to closed systems where critical phenomena arise from
non-analyticities of the Hamiltonian low energy spectrum, in open systems
critical phenomena are intimately related to the low excitation spectrum of
the Liouville operator.  We investigated a generic driven and damped central
spin model and its rich steady state behavior, including critical effects such
as bistabilities, first and second order phase transitions and altered spin pumping dynamics. We
developed a two-step perturbative theory involving the expansion of nuclear
fluctuations up to second order in a self-consistent Holstein-Primakoff transformation and the subsequent adiabatic elimination of
the electron degrees of freedom in the vicinity to the steady state, which
enabled us to provide a complete picture of the system's phase diagram. Linking
common ideas from closed system phase transitions to the dissipative scenario,
we were able to introduce a classification of the different transitions
in the phase diagram.

The relevance of the considered model involves two aspects. On the one hand
\eqref{eq:meq} describes a simple yet rich model, which displays a
large variety of critical phenomena. The limitation to symmetric states allows
for an efficient (and in the thermodynamic limit exact) perturbative treatment
that gives deep insights into the nature of dissipative critical phenomena
from a fundamental point of view.  On the other hand the central spin model is
general enough to have realizations in a large variety of physical systems
(e.g., quantum dots, Nitrogen-Vacancy centers). Our understanding of the critical phenomena in this model could provide insight into recent observation of critical behavior in such systems \cite{Krebs2010,RFTS07}.
Furthermore the main features
of the phase diagram discussed above can also be found if the central
(two-level) spin is replaced by a different physical system, e.g., a larger
spin or a bosonic mode. The theory developed in Section~\ref{sec:PS} can
straightforwardly be adapted to different scenarios and opens the possibility to
study dissipative critical effects in a variety of different physical
systems \cite{Baumann2010a}.

Finally, we showed that in a more realistic adaptation of the model
incorporating an inhomogeneous hyperfine coupling, the second order phase
transition persists, indicating the possibility that the phase diagram remains
qualitatively correct in this experimentally more realistic case. A more
thorough analysis of the effects of inhomogeneities is subject to future work.

\acknowledgements
We acknowledge support by the DFG within SFB 631 and the Cluster of Excellence NIM (E. M. K., G. G. and I. C.), the NSF (M. L. and S. F. Y.), CUA and the Packard Foundation (M. L.), as well as the ECR (A. I.) and AFOSR under the MURI award FA9550-09-1-0588 (S. F. Y.).
\appendix

\section{Approximate Eigenstates of the Lowering Operator}
\subsection{Homogeneous case}
\label{app:exponentialstates}
In Section~\ref{sec:phen} we have seen that we can construct the exact steady state along segment $x$ if we assume the nuclear system to be in an eigenstate of the spin lowering operator
$I^-\ket \alpha = \alpha \ket \alpha$. Although it readily can be show the this operator exactly features only the eigenvalue $\alpha=0$, we can construct approximate eigenvalues in an expansion in $1/J$.

We stress the point that in the bosonic analogue eigenstates of the annihilation operator are coherent minimum uncertainty states that display no squeezing. As we will see, the eigenvectors of the atomic lowering operator in contrast are squeezed coherent atomic states (on the southern hemisphere of the Bloch sphere), where the squeezing parameter depends uniquely on the rotation angle of the Bloch vector. 

As noted in Section~\ref{sec:PS} the Holstein-Primakoff transformation \eqref{eq:HP} provides an exact mapping between  spin operators and a bosonic operator in the subspace of total spin quantum number $J$. In the following we show that approximate eigenstates of the lowering operator $I^-$ can be expressed as a squeezed and displaced vacuum of the bosonic mode $b$
\begin{align}
\label{eq:eigen}
D(\beta) S(-r(\beta))\ket 0 =: \ket{\beta},
\end{align}
where $D(\beta)=e^{\sqrt J \beta b^\dagger -\sqrt J \beta^* b}$ and $S(r)= e^{(r^*b^2 - r b^{\dagger2} )/2}$ are the displacement and squeezing operators, respectively and $\ket 0 \equiv\ket{J-J}$ the fully polarized nuclear state. We find the squeezing parameter uniquely defined by the displacement $r=r(\beta)$. 

Without loss of generality we assume $\beta \in \mathbb R$ (and thus $r\in \mathbb R$), i.e., the Bloch vector lies in the $x-z$ plane. General states $\beta \in \mathbb C$ with arbitrary Bloch vectors on the southern hemisphere, can straightforwardly be derived by a rotation around the $z$-axis. Note that the corresponding states on the northern hemisphere can be constructed accordingly as eigenstates of the ascending operator $I^+$.

In order to show that \eqref{eq:eigen} defines an approximate eigenstate of $I^-$ we first consider the transformation of the nuclear operator under the displacement and squeezing operator.
Recall that according to \eqref{eq:expansion} the displaced nuclear operators can be expanded in orders of $\epsilon=1/\sqrt{J}$
\begin{align}\label{eq:A2}
D^\dagger&(\beta) I^-D(\beta)\\\nonumber
&=
  \sqrt{2J- (\bd + \sqrt{J} \beta^*) (b+ \sqrt{J} \beta)}\left(b + \sqrt{J} \beta \right)\\\nonumber
  &=J \mathcal J_0^- + \sqrt{J} \mathcal J_1^- +\mathcal O(1),
\end{align}
where
\begin{align}
\mathcal J_0^-& = \sqrt k \beta,\\\nonumber
\mathcal J_1^-&= \sqrt{2(1-\beta^2)}(\mu b + \nu b^\dagger)\\\label{eq:r}
                          &= \sqrt{2(1-\beta^2)} S^\dagger(r) b S(r),
 \end{align}
 and $\textrm{cosh}(r)=\mu = \frac{2k - \beta^2}{ 2 \sqrt{2k(1-\beta^2)}}$ and $\textrm{sinh}(r) = - \nu = \frac{\beta^2}{ 2 \sqrt{2k(1-\beta^2)}}$, which defines $r=r(\beta)$ (the generalization to complex $\beta$ is straightforward and leads to \eqref{eq:nanana}). Thus it follows
\begin{align}
S^\dagger(-r)D^\dagger(\beta) &I^-D(\beta)S(-r)\ket 0\\\nonumber
&= J \mathcal J_0^- \ket 0 + \mathcal O(1),
\end{align}
since $b\ket 0 =0$.

Multiplying both sides by $D(\beta)S(-r)$ yields the desired approximate eigenvalue equation
\begin{align}
&  I^-\ket{\beta} =J \sqrt k \beta \ket{\beta} + \mathcal O(1).\label{eq:eq}
\end{align}
In the thermodynamic limit the term $\mathcal O (1)$ is negligible and the eigenvalue equation is exact \footnote{This is true even for $\beta \rightarrow 0 $ since all terms in the expansion \eqref{eq:A2} that do not vanish upon application on $\ket 0$ contain at least one factor $\beta$ as well.}.


Using the above representation we study the spin properties of the states $\ket \alpha$.
In the following all expectation values  are understood to be evaluated in the squeezed coherent state $\ket\beta$: $\E O \equiv \bra\beta O \ket\beta$.

Straightforwardly, one derives the nuclear mean polarizations
\begin{align}
\E {I_x} &= \frac{1}{2}\bra \beta (I^+ + I^-) \ket \beta = J\sqrt k\beta + \mathcal O(1),\\
\E {I_y} &= \frac{1}{2i}\bra \beta (I^+ - I^-) \ket \beta = 0 + \mathcal O(1),\\
\E {I_z} &=  J(\beta^2 - 1) + \mathcal O(1),
\end{align}
where in the last equation we used the expansion \eqref{eq:Jz}. Note that the Bloch vector is orthogonal (up to order $\mathcal O(1)$) to the $y$-direction for all (real) $\alpha$ and of length $|\E{\vec{I}}|=\sqrt{\E {I_x}^2 + \E {I_y}^2 + \E {I_z}^2} =J + \mathcal O(1)$.

Using \eqref{eq:eq} and the angular momentum commutation relations one readily calculates
\begin{align}
\E {\Delta I_y^2} &=- \frac{1}{2} \E {I_z}  + \mathcal O(1),\\\nonumber
                              &=  \frac{1}{2} J(1-\beta^2 ) + \mathcal O(1),\\\nonumber
                              &=  \frac{1}{2}J \sqrt{1-(\sqrt k\beta)^2} + \mathcal O(1)
\end{align}
where as usual $\E {\Delta O ^2} := \E {O^2} - \E O^2$ and we used the identity $1-(\sqrt k \beta)^2 = (1-\beta^2)^2$.

Thus, we find for the squeezing parameter in $y$-direction,
\begin{align}\label{eq:squeezingapp}
\xi_y^2=2\E{\Delta I_y^2}/|\E{\vec I}|=\sqrt{1-(\sqrt k\beta)^2} + \mathcal O(1/J).
\end{align}
The squeezing diverges for the state that realizes the maximal eigenvalue of the lowering operator ($\sqrt k\beta = 1$). This corresponds to a state fully polarized in $x$ direction.

\subsection{Inhomogeneous case}
\label{app:inhomo}

We approximate a system of inhomogeneous hyperfine coupling by grouping the nuclear spins into n shells. Within a shell $i$ the nuclear spins have identical coupling $g_i$ and the respective (homogeneous) spin operators $A^{(i)}_\alpha$ ($\alpha=x,y,z$) are truncated to a symmetric subspace $J_i$. The total spin operators can then be written as
\begin{align}
A_\alpha=\sum_{i=1}^n g_i A^{(i)}_\alpha.
\end{align}
We define collective displacement and squeezing operators
\begin{align}
\mathcal D= \Pi_{i=1}^n e^{\sqrt J_i \beta_i b_i^\dagger -\sqrt J_i \beta_i^* b_i},\\
\mathcal S= \Pi_{i=1}^n  e^{(r_i^*b_i^2 - r_i b_i^{\dagger2} )/2},
\end{align}
where the $b_i$ is the respective bosonic operator for shell $i$. Also here the squeezing parameter $r_i$ depends uniquely (with the same functional dependence as before, cf. \eqref{eq:r}) on the displacement $\beta_i$ within the shell, if we demand the first order in the eigenvalue equation to vanish:
\begin{align}
&  A^- \mathcal D \mathcal S\ket{0} =(\sum_i J_i \sqrt k_i \beta_i) \mathcal D \mathcal S\ket{0} + \mathcal O(1),
\end{align}
where $k_i=\sqrt{2-\beta_i^2}$ and $\ket0 \equiv \ket 0^{\otimes n}$ is the vacuum of the shell modes.

We emphasize that in general the eigenvalues are highly degenerate. For a given eigenvalue $\alpha$ there are infinitely many microscopic realizations (i.e. sets of $\beta_i$) that fulfill $\alpha = \sum_i J_i \sqrt k_i \beta_i$. Only the maximal eigenvalue $\alpha=J$ features a unique steady state that displays diverging squeezing as one readily shows analogous to the homogeneous case.

\section{Rotated Squeezed Thermal Spin States}
\label{app:RSTSS}

A key concept of the paper are RSTSS, a generalization of squeezed coherent spin states to mixed states, parametrized via an effective temperature. They describe nuclear states which are fully polarized, rotated, and feature fluctuations which can be described by a bosonic mode in a thermal (potentially squeezed) Gaussian state. 
In Section~\ref{sec:theory} we show that the truncation of every nuclear operator to a subspace of total spin $J$ can be expressed in terms of a bosonic mode $b$ and its displacement $\beta \in \mathbb C$, using a Holstein-Primakoff transformation [compare \eqref{eq:expansion}, \eqref{eq:Jz0}]
\begin{align}\label{eq:appexp}
I^\alpha / J  = \sum_n \epsilon^n \mathcal J_n^\alpha,
\end{align}
where $\epsilon=1/\sqrt J$, and the bosonic operators $\mathcal J_n^\alpha$ contain combinations of products of $n$ bosonic operators $b, b^\dagger$.
$ \mathcal J_0^\alpha \in \mathbb C$, describes the semiclassical expectation value which is fully determined by the displacement $\beta$. $\beta$ quantifies a rotation of the fully polarized nuclear state on the Bloch sphere.  
The higher order operators $ \mathcal J_n^\alpha$ ($n>0$) describe quantum fluctuations around this semiclassical nuclear state. RSTSS are those states where the mode $b$ is in an undisplaced ($\E{b}=0$), squeezed thermal state, which is fully determined by its covariance matrix $\Gamma$ [\eqref{eq:covmat1}]. These bosonic states constitute the natural steady states of the quadratic Master \eqref{eq:effmeq}, and we find in Section~\ref{sec:ptgm} that across the whole phase diagram one steady state of the system can always be described as a RSTSS.

Note that in the limit where the effective temperature of the Gaussian state is zero, we recover the class of squeezed coherent spin states \cite{Kitagawa:1993di}, which constitute the solution along segment $x$. 


\section{Solving \eqref{eq:beta}}
\label{app:beta}

In order to find the solutions to \eqref{eq:beta} (which are numerically difficult to find) we first note that
\begin{align}
\E{A}_{ss}=0 \Leftrightarrow \E{\dot b} = \E{ \dot\bd }= 0 \Leftrightarrow \E{ \dot{\mathcal{J}_1^-}} =  \E{ \dot{\mathcal{J}_1^+}}=0 ,
\end{align}
where the time derivative is understood with respect to the first order Liouvillian
\begin{align}
\mathcal{L}_{1} \rho= &- i [a (S_x  \mathcal J^x_{1} + S_y\mathcal J^y_{1}) + (a S^+S^-  + \delta\omega) \mathcal J_{1}^z,\rho],
\end{align}
and in the usual way we define
\begin{align}
 \mathcal J^x_{1}= \frac{1}{2}(\mathcal J^+_{1} + \mathcal J^-_{1}),\\
 \mathcal J^y_{1}= \frac{1}{2i}(\mathcal J^+_{1} - \mathcal J^-_{1}).
\end{align}
Using the relation $\left[ J^i_{1}, J^j_{1} \right]=i \epsilon_{ijk} J^k_{0}$ one finds the equations
\begin{align}
0=\E{ \dot{\mathcal{J}_1^x}} &=a\left(\E {S_y}_{ss} \mathcal J^z_{0}  - \E {S_z}_{ss}\mathcal J^y_{0}  \right)-\omega\mathcal J^y_{0} ,\\
0=\E{ \dot{\mathcal{J}_1^y}} &= -a\left(\E {S_x}_{ss}\mathcal J^z_{0}  - \E {S_z}_{ss}\mathcal J^x_{0}  \right)+\omega\mathcal J^x_{0} ,\\
0=\E{ \dot{\mathcal{J}_1^z}} &=a\left(\E {S_y}_{ss}\mathcal J^x_{0}  - \E {S_x}_{ss}\mathcal J^y_{0}  \right).
\end{align}
Furthermore from the definitions of the $\mathcal J^i_{0}$'s one finds
\begin{align}
1=(\mathcal J^x_{0})^2 + (\mathcal J^y_{0})^2 +(\mathcal J^z_{0})^2.
\end{align}

The steady state expectation values $\E{S_i}_{ss}$ are found directly via [cf. \eqref{eq:L0}]
\begin{align}
\mathcal{L}_0 \rho=& \gamma(S^-\rho S^+ - \frac{1}{2} \{ S^+S^-,\rho \}_+) \\\nonumber
&- i [S_x(2 \Omega + a \mathcal J^x_0) +a S_y  \mathcal J^y_0 + a S^+S^- \mathcal J^z_0,\rho],
\end{align}
by solving the resulting optical Bloch equations
\begin{align}
0& = -\frac{\gamma}{2} \E{S_x} + a  \mathcal J^y_0 \E{S_z} - a \mathcal J^z_0 \E{S_y},\\
0 &=  -\frac{\gamma}{2} \E{S_y} - (2\Omega +  a  \mathcal J^x_0) \E{S_z} + a \mathcal J^z_0 \E{S_x},\\
0&=  -\gamma (\E{S_z}+1/2) + (2\Omega +  a  \mathcal J^x_0) \E{S_y} - a \mathcal J^y_0 \E{S_x}.
\end{align}
This set of coupled Bloch equations for the six variables $\{ \E{S_i}, \mathcal J^j_0   \}$ can be solved analytically. The solutions which feature second order stability (see Section~\ref{sec:methods}) are displayed in \figref{fig:1}.
Via \eqref{eq:Jp} and \eqref{eq:Jz} $\beta$ can be deduced unambiguously from a given set $\{ \E{S_i}, \mathcal J^j_0   \}$.

\section{Deriving the second order term of \eqref{eq:meqeff}}
\label{app:2ndorder}

The first term of the second order of \eqref{eq:meqeff} is of the same form as the first order and can readily be calculated:
\begin{align}
\label{eq:fo}
Tr_S(P\mathcal L_2 P\rho) =- i [&a/2 (\E{S^+}_{ss}  \mathcal J^-_{2} + \E{S^-}_{ss}\mathcal J^+_{2})   \\\nonumber
&+ (a \E{S^+S^-}_{ss}+ \delta\omega) \mathcal J_{2}^z,\sigma ],\\\nonumber
=-i [&B^* b^2 + B (\bd)^2 + F \bd b,\sigma],
\end{align}
with the $\beta$-dependent coefficients (remember that also the electron steady state expectation values are functions of $\beta$)
\begin{align}
\label{eq:coeff1a}
B=& -\frac{a\beta }{16 \sqrt{k^3} }  \left[ ( 4k + |\beta|^2   )  \E{ S^-}_{ss} + \beta^2  \E{ S^+}_{ss}   \right],\\
\label{eq:coeff11a}
F=& -\frac{a}{8 \sqrt{k^3} }   ( 4k + |\beta|^2) \left(\beta \E{S^+}_{ss} + \beta^* \E{S^-}_{ss} \right) \\\nonumber
&+ a (\E{S^+S^-}_{ss} + \delta\omega/a).
\end{align}


Next, we consider the second term of the second order perturbative master equation
\begin{align}
\label{eq:2ndorder}
-Tr_s&( P \mathcal L_1 Q \mathcal L_0^{-1} Q\mathcal L_1P\rho) \\\nonumber
=& -Tr_s( P \mathcal L_1 (\mathbb1 -P) \mathcal L_0^{-1} (\mathbb 1 -P) \mathcal L_1P\rho)\\\nonumber
 =& \int_0^\mathcal1 d\tau Tr_s  (P \mathcal L_1 e^{ \mathcal L_0 \tau} \mathcal L_1P\rho) \\\nonumber
 &-  \int_0^\mathcal1 d\tau Tr_s  (P \mathcal L_1 P \mathcal L_1P\rho),
\end{align}
where we used the Laplace transform $-\mathcal L_0^{-1} =  \int_0^\mathcal1 d\tau  e^{ \mathcal L_0 \tau}$ and the property $e^{ \mathcal L_0 \tau}P=Pe^{ \mathcal L_0 \tau}=P$.

Noting that
\begin{align}
Tr_s( P\mathcal L_1 X) =-i Tr_s(\left[b A + \bd A^\dagger , X   \right]),
\end{align}
and using \eqref{eq:1} we find
\begin{align}
-\int_0^\mathcal1 &d\tau Tr_s  (P \mathcal L_1 P \mathcal L_1P\rho)\\\nonumber
 =&  \int_0^\mathcal1 d\tau \E{A^\alpha}_{ss}  \E{A^\beta}_{ss} \left[b^\alpha, \left[b^\beta, \sigma \right] \right] ,
\end{align}
where $\alpha,\beta = \dagger, \mathrm{'void'}$ and the Einstein sum convention is used.

In the same fashion we find
\begin{align}
 \int_0^\mathcal1 &d\tau Tr_s  (P \mathcal L_1 e^{ \mathcal L_0 \tau} \mathcal L_1P\rho)\\\nonumber
 =& - \int_0^\mathcal1 d\tau \E{A^\alpha(\tau) A^\beta(0)}_{ss} \left[b^\alpha, \left[b^\beta, \sigma \right] \right]\\\nonumber
 &- \int_0^\mathcal1 d\tau \E{\left[A^\alpha(\tau), A^\beta(0)\right]}_{ss} \left[b^\alpha, \sigma b^\beta \right].
\end{align}
Here we defined the autocorrelation functions $\E{A^\alpha(\tau) A^\beta(0)}_{ss} = Tr_s(A^\alpha e^{ \mathcal L_0 \tau} A^\beta \rho_{ss})$ and $\E{\left[A^\alpha(\tau), A^\beta(0)\right]}_{ss} = Tr_s(A^\alpha e^{ \mathcal L_0 \tau} \left[A^\beta, \rho_{ss}\right])$ (cf. e.g., \cite{Carmichael1999} pp. 22).

Putting together the results \eqref{eq:2ndorder} reduces to
\begin{align}
\label{eq:2ndorder2}
-Tr_s&( P \mathcal L_1 Q \mathcal L_0^{-1} Q\mathcal L_1P\rho) \\\nonumber = &  - \int_0^\mathcal1 d\tau \E{\Delta A^\alpha(\tau)\Delta A^\beta(0)}_{ss} \left[b^\alpha, \left[b^\beta, \sigma \right] \right]\\\nonumber
 &- \int_0^\mathcal1 d\tau \E{\left[\Delta A^\alpha(\tau),\Delta  A^\beta(0)\right]}_{ss} \left[b^\alpha, \sigma b^\beta \right],
\end{align}
$\Delta O := O - \E{O}_{ss} $. Since we choose the displacement $\beta$ such that $\E{A^\alpha}_{ss}= 0$ [\eqref{eq:beta}] it is $\Delta A^\alpha = A^\alpha $. Merging \eqref{eq:fo} and \eqref{eq:2ndorder2}, and regrouping the terms, one readily derives equation \eqref{eq:effmeq}.

\subsection{Calculation of the coefficients}
\label{app:coeff}

In order to determine the coefficients \eqref{eq:coeff2} we have to calculate terms of the kind $\int_0^\mathcal 1 d\tau \E{\Delta A^\alpha(\tau)\Delta A^\beta(0)}_{ss}$ and $\int_0^\mathcal 1 d\tau \E{\Delta A^\alpha(0)\Delta A^\beta(\tau)}_{ss}$. Exemplarily we will calculate the two terms for $\alpha=\beta=\mathrm{'void'}$.

First, defining $\vec v=(\frac{a}{4\sqrt{k}} (2k-|\beta|^2), -\frac{a}{4\sqrt{k}}\beta^2 , \beta a)^T$ we can write $\Delta A = \vec v^* \cdot \Delta \vec S$ (and with $\vec w=( -\frac{a}{4\sqrt{k}}(\beta^*)^2,\frac{a}{4\sqrt{k}} (2k-|\beta|^2), \beta^* a)^T $ we find  $\Delta A^\dagger = \vec w^* \cdot \Delta \vec S $ ). Likewise it is $\Delta A^\dagger = \Delta \vec S^\dagger \cdot   \vec v$ ($\Delta A = \Delta \vec S^\dagger \cdot   \vec w$).

Consequently we compute:
\begin{align}
\int_0^\mathcal 1 &d\tau \E{\Delta A_\tau \Delta A }_{ss}\\\nonumber
= &\vec{v}^* \left(\int_0^\mathcal 1 d\tau \E{\Delta \vec S_\tau \Delta \vec S^\dagger}_{ss}\right) \vec w\\\nonumber
 =& \vec{v}^* \left(\int_0^\mathcal 1 d\tau e^{\mathcal M \tau} \E{\Delta \vec S  \Delta \vec S^\dagger}_{ss}\right) \vec w \\\nonumber
=&\vec{v}^* \left( - \mathcal M ^{-1}\E{\Delta \vec S  \Delta \vec S^\dagger}_{ss}\right) \vec w = \vec{v}^*  \mathcal F_1 \vec w,
\end{align}
where we applied the Quantum Regression Theorem in the second step and used the definitions of
Section~(\ref{sec:theory}).

Noting that
\begin{align}
\int_0^{\mathcal 1} &d\tau \E{\Delta \vec S \Delta \vec S _\tau^\dagger}_{ss}= \left( \int_0^{\mathcal 1} d\tau \E{\Delta \vec S_\tau \Delta \vec S^\dagger}_{ss} \right)^\dagger \\\nonumber
=& \left( - \mathcal M ^{-1}\E{\Delta \vec S  \Delta \vec S^\dagger}_{ss} \right)^\dagger \\\nonumber
=&  - \E{\Delta \vec S  \Delta \vec S^\dagger}_{ss} \mathcal M ^{-\dagger} = \mathcal F_2 = \mathcal F_1^\dagger,
\end{align}
we write
\begin{align}
\int_0^\mathcal 1 d\tau \E{\Delta A \Delta A_\tau }_{ss} = \vec{v}^* \mathcal F_2 \vec w.
\end{align}

Analogously, we find the relations
\begin{align}
\int_0^\mathcal 1 d\tau \E{\Delta A^\dagger_\tau \Delta A }_{ss} &= \vec{w}^* \mathcal F_1 \vec w, \\\nonumber
\int_0^\mathcal 1 d\tau \E{\Delta A^\dagger \Delta A_\tau }_{ss} &= \vec{w}^* \mathcal F_2 \vec w, \\\nonumber
\int_0^\mathcal 1 d\tau \E{\Delta A_\tau \Delta A^\dagger }_{ss} &= \vec{v}^* \mathcal F_1 \vec v, \\\nonumber
\int_0^\mathcal 1 d\tau \E{\Delta A \Delta A_\tau^\dagger }_{ss} &= \vec{v}^* \mathcal F_2 \vec v, \\\nonumber
&\vdots
\end{align}
such that all coefficients of the effective Master \eqref{eq:meqeff} can be calculated by simple matrix multiplication.

%
%
%

%
%







\begin{thebibliography}{49}
\expandafter\ifx\csname natexlab\endcsname\relax\def\natexlab#1{#1}\fi
\expandafter\ifx\csname bibnamefont\endcsname\relax
  \def\bibnamefont#1{#1}\fi
\expandafter\ifx\csname bibfnamefont\endcsname\relax
  \def\bibfnamefont#1{#1}\fi
\expandafter\ifx\csname citenamefont\endcsname\relax
  \def\citenamefont#1{#1}\fi
\expandafter\ifx\csname url\endcsname\relax
  \def\url#1{\texttt{#1}}\fi
\expandafter\ifx\csname urlprefix\endcsname\relax\def\urlprefix{URL }\fi
\providecommand{\bibinfo}[2]{#2}
\providecommand{\eprint}[2][]{\url{#2}}

\bibitem[{\citenamefont{Vojta}(2003)}]{Vojta2003}
\bibinfo{author}{\bibfnamefont{M.}~\bibnamefont{Vojta}},
  \bibinfo{journal}{Reports on Progress in Physics}
  \textbf{\bibinfo{volume}{66}}, \bibinfo{pages}{2069} (\bibinfo{year}{2003}).

\bibitem[{\citenamefont{Sachdev}(2003)}]{Sachdev:2003fp}
\bibinfo{author}{\bibfnamefont{S.}~\bibnamefont{Sachdev}},
  \bibinfo{journal}{Reviews of Modern Physics} \textbf{\bibinfo{volume}{75}},
  \bibinfo{pages}{913} (\bibinfo{year}{2003}).

\bibitem[{\citenamefont{Ginzburg}(2007)}]{Ginzburg:2007dz}
\bibinfo{author}{\bibfnamefont{V.~L.} \bibnamefont{Ginzburg}},
  \bibinfo{journal}{Physics-Uspekhi} \textbf{\bibinfo{volume}{40}},
  \bibinfo{pages}{407} (\bibinfo{year}{2007}).

\bibitem[{\citenamefont{Kim and Chan}(2004)}]{Kim:2004hl}
\bibinfo{author}{\bibfnamefont{E.}~\bibnamefont{Kim}} \bibnamefont{and}
  \bibinfo{author}{\bibfnamefont{M.~H.~W.} \bibnamefont{Chan}},
  \bibinfo{journal}{Nature} \textbf{\bibinfo{volume}{427}},
  \bibinfo{pages}{225} (\bibinfo{year}{2004}).

\bibitem[{\citenamefont{Belitz and Kirkpatrick}(1994)}]{Belitz:1994is}
\bibinfo{author}{\bibfnamefont{D.}~\bibnamefont{Belitz}} \bibnamefont{and}
  \bibinfo{author}{\bibfnamefont{T.}~\bibnamefont{Kirkpatrick}},
  \bibinfo{journal}{Reviews of Modern Physics} \textbf{\bibinfo{volume}{66}},
  \bibinfo{pages}{261} (\bibinfo{year}{1994}).

\bibitem[{\citenamefont{Hasan and Kane}(2010)}]{Hasan2010}
\bibinfo{author}{\bibfnamefont{M.}~\bibnamefont{Hasan}} \bibnamefont{and}
  \bibinfo{author}{\bibfnamefont{C.}~\bibnamefont{Kane}},
  \bibinfo{journal}{Reviews of Modern Physics} \textbf{\bibinfo{volume}{82}},
  \bibinfo{pages}{3045} (\bibinfo{year}{2010}).

\bibitem[{\citenamefont{Carmichael}(1980)}]{HJCarmichael1980}
\bibinfo{author}{\bibfnamefont{H.~J.} \bibnamefont{Carmichael}},
  \bibinfo{journal}{Journal of Physics B: Atomic and Molecular Physics}
  \textbf{\bibinfo{volume}{13}}, \bibinfo{pages}{3551} (\bibinfo{year}{1980}).

\bibitem[{\citenamefont{Werner et~al.}(2005)\citenamefont{Werner, Volker,
  Troyer, and Chakravarty}}]{Werner2005}
\bibinfo{author}{\bibfnamefont{P.}~\bibnamefont{Werner}},
  \bibinfo{author}{\bibfnamefont{K.}~\bibnamefont{Volker}},
  \bibinfo{author}{\bibfnamefont{M.}~\bibnamefont{Troyer}}, \bibnamefont{and}
  \bibinfo{author}{\bibfnamefont{S.}~\bibnamefont{Chakravarty}},
  \bibinfo{journal}{Physical Review Letters} \textbf{\bibinfo{volume}{94}},
  \bibinfo{pages}{047201} (\bibinfo{year}{2005}).

\bibitem[{\citenamefont{Capriotti et~al.}(2005)\citenamefont{Capriotti,
  Cuccoli, Fubini, Tognetti, and Vaia}}]{Capriotti2005}
\bibinfo{author}{\bibfnamefont{L.}~\bibnamefont{Capriotti}},
  \bibinfo{author}{\bibfnamefont{A.}~\bibnamefont{Cuccoli}},
  \bibinfo{author}{\bibfnamefont{A.}~\bibnamefont{Fubini}},
  \bibinfo{author}{\bibfnamefont{V.}~\bibnamefont{Tognetti}}, \bibnamefont{and}
  \bibinfo{author}{\bibfnamefont{R.}~\bibnamefont{Vaia}},
  \bibinfo{journal}{Physical Review Letters} \textbf{\bibinfo{volume}{94}},
  \bibinfo{pages}{157001} (\bibinfo{year}{2005}).

\bibitem[{\citenamefont{Morrison and Parkins}(2008)}]{Morrison2008b}
\bibinfo{author}{\bibfnamefont{S.}~\bibnamefont{Morrison}} \bibnamefont{and}
  \bibinfo{author}{\bibfnamefont{A.~S.} \bibnamefont{Parkins}},
  \bibinfo{journal}{Journal of Physics B: Atomic, Molecular and Optical
  Physics} \textbf{\bibinfo{volume}{41}}, \bibinfo{pages}{195502}
  (\bibinfo{year}{2008}).

\bibitem[{\citenamefont{Eisert and Prosen}(2010)}]{Eisert:2010vj}
\bibinfo{author}{\bibfnamefont{J.}~\bibnamefont{Eisert}} \bibnamefont{and}
  \bibinfo{author}{\bibfnamefont{T.}~\bibnamefont{Prosen}},
  \bibinfo{journal}{arXiv:1012.5013}
  (\bibinfo{year}{2010}).

\bibitem[{\citenamefont{Bhaseen et~al.}(2012)\citenamefont{Bhaseen, Mayoh,
  Simons, and Keeling}}]{Bhaseen12}
\bibinfo{author}{\bibfnamefont{M.~J.} \bibnamefont{Bhaseen}},
  \bibinfo{author}{\bibfnamefont{J.}~\bibnamefont{Mayoh}},
  \bibinfo{author}{\bibfnamefont{B.~D.} \bibnamefont{Simons}},
  \bibnamefont{and} \bibinfo{author}{\bibfnamefont{J.}~\bibnamefont{Keeling}},
  \bibinfo{journal}{Phys. Rev. A} \textbf{\bibinfo{volume}{85}},
  \bibinfo{pages}{013817} (\bibinfo{year}{2012}).

\bibitem[{\citenamefont{Baumann et~al.}(2010)\citenamefont{Baumann, Guerlin,
  Brennecke, and Esslinger}}]{Baumann2010a}
\bibinfo{author}{\bibfnamefont{K.}~\bibnamefont{Baumann}},
  \bibinfo{author}{\bibfnamefont{C.}~\bibnamefont{Guerlin}},
  \bibinfo{author}{\bibfnamefont{F.}~\bibnamefont{Brennecke}},
  \bibnamefont{and}
  \bibinfo{author}{\bibfnamefont{T.}~\bibnamefont{Esslinger}},
  \bibinfo{journal}{Nature} \textbf{\bibinfo{volume}{464}},
  \bibinfo{pages}{1301} (\bibinfo{year}{2010}).
  
 \bibitem[{\citenamefont{\"Oztop et~al.}(2010)}]{Oztop:2011tz}
\bibinfo{author}{\bibfnamefont{B.}~\bibnamefont{\"Oztop}},
  \bibinfo{author}{\bibfnamefont{M.}~\bibnamefont{Bordyuh}},
  \bibinfo{author}{\bibfnamefont{\"O. E.}~\bibnamefont{M\"ustecapl\i o\u glu}},
  \bibnamefont{and}
  \bibinfo{author}{\bibfnamefont{H. E.}~\bibnamefont{T\"ureci}},
  \bibinfo{journal}{arxiv:1107.3108} (\bibinfo{year}{2011}).
  

\bibitem[{\citenamefont{Hepp and Lieb}(1973)}]{Hepp1973}
\bibinfo{author}{\bibfnamefont{K.}~\bibnamefont{Hepp}} \bibnamefont{and}
  \bibinfo{author}{\bibfnamefont{E.~H.} \bibnamefont{Lieb}},
  \bibinfo{journal}{Annals of Physics} \textbf{\bibinfo{volume}{76}},
  \bibinfo{pages}{360} (\bibinfo{year}{1973}).

\bibitem[{\citenamefont{Gibbs et~al.}(1976)\citenamefont{Gibbs, McCall, and
  Venkatesan}}]{Gibbs1976}
\bibinfo{author}{\bibfnamefont{H.}~\bibnamefont{Gibbs}},
  \bibinfo{author}{\bibfnamefont{S.}~\bibnamefont{McCall}}, \bibnamefont{and}
  \bibinfo{author}{\bibfnamefont{T.}~\bibnamefont{Venkatesan}},
  \bibinfo{journal}{Physical Review Letters} \textbf{\bibinfo{volume}{36}},
  \bibinfo{pages}{1135} (\bibinfo{year}{1976}).

\bibitem[{\citenamefont{Lawande et~al.}(1981)\citenamefont{Lawande, Puri, and
  Hassan}}]{Lawande1981}
\bibinfo{author}{\bibfnamefont{S.~V.} \bibnamefont{Lawande}},
  \bibinfo{author}{\bibfnamefont{R.~R.} \bibnamefont{Puri}}, \bibnamefont{and}
  \bibinfo{author}{\bibfnamefont{S.~S.} \bibnamefont{Hassan}},
  \bibinfo{journal}{Journal of Physics B: Atomic and Molecular Physics}
  \textbf{\bibinfo{volume}{14}}, \bibinfo{pages}{4171} (\bibinfo{year}{1981}).

\bibitem[{\citenamefont{Puri et~al.}(1980)\citenamefont{Puri, Lawande, and
  Hassan}}]{Puri1980}
\bibinfo{author}{\bibfnamefont{R.~R.} \bibnamefont{Puri}},
  \bibinfo{author}{\bibfnamefont{S.~V.} \bibnamefont{Lawande}},
  \bibnamefont{and} \bibinfo{author}{\bibfnamefont{S.~S.}
  \bibnamefont{Hassan}}, \bibinfo{journal}{Optics Communications}
  \textbf{\bibinfo{volume}{35}}, \bibinfo{pages}{179} (\bibinfo{year}{1980}).

\bibitem[{\citenamefont{Bonifacio and Lugiato}(1978)}]{Bonifacio1978}
\bibinfo{author}{\bibfnamefont{R.}~\bibnamefont{Bonifacio}} \bibnamefont{and}
  \bibinfo{author}{\bibfnamefont{L. A.}~\bibnamefont{Lugiato}},
  \bibinfo{journal}{Physical Review A} \textbf{\bibinfo{volume}{18}},
  \bibinfo{pages}{1129} (\bibinfo{year}{1978}).

\bibitem[{\citenamefont{Schneider and Milburn}(2002)}]{Schneider2002a}
\bibinfo{author}{\bibfnamefont{S.}~\bibnamefont{Schneider}} \bibnamefont{and}
  \bibinfo{author}{\bibfnamefont{G. J.}~\bibnamefont{Milburn}},
  \bibinfo{journal}{Physical Review A} \textbf{\bibinfo{volume}{65}},
  \bibinfo{pages}{042107} (\bibinfo{year}{2002}).

\bibitem[{\citenamefont{Gaudin}(1976)}]{Gaudin1976}
\bibinfo{author}{\bibfnamefont{M.}~\bibnamefont{Gaudin}}, \bibinfo{journal}{J.
  Phys. France} \textbf{\bibinfo{volume}{37}}, \bibinfo{pages}{1087}
  (\bibinfo{year}{1976}).

\bibitem[{\citenamefont{Bortz and Stolze}(2007)}]{BorSt07}
\bibinfo{author}{\bibfnamefont{M.}~\bibnamefont{Bortz}} \bibnamefont{and}
  \bibinfo{author}{\bibfnamefont{J.}~\bibnamefont{Stolze}},
  \bibinfo{journal}{J. Stat. Mech.} \textbf{\bibinfo{volume}{2007}},
  \bibinfo{pages}{P06018} (\bibinfo{year}{2007}).

\bibitem[{\citenamefont{Schliemann et~al.}(2003)\citenamefont{Schliemann,
  Khaetskii, and Loss}}]{SKL03}
\bibinfo{author}{\bibfnamefont{J.}~\bibnamefont{Schliemann}},
  \bibinfo{author}{\bibfnamefont{A.}~\bibnamefont{Khaetskii}},
  \bibnamefont{and} \bibinfo{author}{\bibfnamefont{D.}~\bibnamefont{Loss}},
  \bibinfo{journal}{J. Phys: Cond. Mat.} \textbf{\bibinfo{volume}{15}},
  \bibinfo{pages}{R1809} (\bibinfo{year}{2003}).

\bibitem[{\citenamefont{Krebs et~al.}(2010)\citenamefont{Krebs, Maletinsky,
  Amand, Urbaszek, Lema{\^\i}tre, Voisin, Marie, and Imamoglu}}]{Krebs2010}
\bibinfo{author}{\bibfnamefont{O.}~\bibnamefont{Krebs}},
  \bibinfo{author}{\bibfnamefont{P.}~\bibnamefont{Maletinsky}},
  \bibinfo{author}{\bibfnamefont{T.}~\bibnamefont{Amand}},
  \bibinfo{author}{\bibfnamefont{B.}~\bibnamefont{Urbaszek}},
  \bibinfo{author}{\bibfnamefont{A.}~\bibnamefont{Lemaitre}},
  \bibinfo{author}{\bibfnamefont{P.}~\bibnamefont{Voisin}},
  \bibinfo{author}{\bibfnamefont{X.}~\bibnamefont{Marie}}, \bibnamefont{and}
  \bibinfo{author}{\bibfnamefont{A.}~\bibnamefont{Imamoglu}},
  \bibinfo{journal}{Physical Review Letters} \textbf{\bibinfo{volume}{104}},
  \bibinfo{pages}{056603} (\bibinfo{year}{2010}).

\bibitem[{\citenamefont{Russell et~al.}(2007)\citenamefont{Russell, Fal'ko,
  Tartakovskii, and Skolnick}}]{RFTS07}
\bibinfo{author}{\bibfnamefont{A.}~\bibnamefont{Russell}},
  \bibinfo{author}{\bibfnamefont{V.~I.} \bibnamefont{Fal'ko}},
  \bibinfo{author}{\bibfnamefont{A.~I.} \bibnamefont{Tartakovskii}},
  \bibnamefont{and} \bibinfo{author}{\bibfnamefont{M.~S.}
  \bibnamefont{Skolnick}}, \textbf{\bibinfo{volume}{76}},
  \bibinfo{pages}{195310} (\bibinfo{year}{2007}).

\bibitem[{Riv(2012)}]{Rivas:2012jr}
\bibinfo{author}{\bibfnamefont{A.}~\bibnamefont{Rivas}},
\bibinfo{author}{\bibfnamefont{S. F.}~\bibnamefont{Huelga}},
  \emph{\bibinfo{title}{{Open Quantum Systems: An Introduction}}}
  (\bibinfo{publisher}{Springer},
  \bibinfo{address}{Berlin}, \bibinfo{year}{1999}).

\bibitem[{\citenamefont{Horstmann}(2011)}]{Horstmann:oyrhcBUx}
B. Horstmann et al., to be published (2012).

\bibitem[{\citenamefont{Sachdev}(1999)}]{Sachdev2011}
\bibinfo{author}{\bibfnamefont{S.}~\bibnamefont{Sachdev}},
  \emph{\bibinfo{title}{{Quantum Phase Transitions}}}
  (\bibinfo{publisher}{Cambridge University Press},
  \bibinfo{address}{Cambridge}, \bibinfo{year}{1999}).

\bibitem[{\citenamefont{Atat{\"u}re et~al.}(2006)\citenamefont{Atat{\"u}re,
  Dreiser, Badolato, H{\"o}gele, Karrai, and Imamoglu}}]{Atature:2006ha}
\bibinfo{author}{\bibfnamefont{M.}~\bibnamefont{Atat{\"u}re}},
  \bibinfo{author}{\bibfnamefont{J.}~\bibnamefont{Dreiser}},
  \bibinfo{author}{\bibfnamefont{A.}~\bibnamefont{Badolato}},
  \bibinfo{author}{\bibfnamefont{A.}~\bibnamefont{H{\"o}gele}},
  \bibinfo{author}{\bibfnamefont{K.}~\bibnamefont{Karrai}}, \bibnamefont{and}
  \bibinfo{author}{\bibfnamefont{A.}~\bibnamefont{Imamoglu}},
  \bibinfo{journal}{Science} \textbf{\bibinfo{volume}{312}},
  \bibinfo{pages}{551} (\bibinfo{year}{2006}).

\bibitem[{\citenamefont{Tamarat et~al.}(2008)\citenamefont{Tamarat, Manson,
  Harrison, McMurtrie, Nizovtsev, Santori, Beausoleil, Neumann, Gaebel, Jelezko
  et~al.}}]{Tamarat:2008ko}
\bibinfo{author}{\bibfnamefont{P.}~\bibnamefont{Tamarat}},
  \bibinfo{author}{\bibfnamefont{N.~B.} \bibnamefont{Manson}},
  \bibinfo{author}{\bibfnamefont{J.~P.} \bibnamefont{Harrison}},
  \bibinfo{author}{\bibfnamefont{R.~L.} \bibnamefont{McMurtrie}},
  \bibinfo{author}{\bibfnamefont{A.}~\bibnamefont{Nizovtsev}},
  \bibinfo{author}{\bibfnamefont{C.}~\bibnamefont{Santori}},
  \bibinfo{author}{\bibfnamefont{R.~G.} \bibnamefont{Beausoleil}},
  \bibinfo{author}{\bibfnamefont{P.}~\bibnamefont{Neumann}},
  \bibinfo{author}{\bibfnamefont{T.}~\bibnamefont{Gaebel}},
  \bibinfo{author}{\bibfnamefont{F.}~\bibnamefont{Jelezko}},
  \bibnamefont{et~al.}, \bibinfo{journal}{New Journal of Physics}
  \textbf{\bibinfo{volume}{10}}, \bibinfo{pages}{045004}
  (\bibinfo{year}{2008}).

\bibitem[{\citenamefont{Kessler et~al.}(2010)\citenamefont{Kessler, Yelin,
  Lukin, Cirac, and Giedke}}]{Kessler:2010fb}
\bibinfo{author}{\bibfnamefont{E.~M.} \bibnamefont{Kessler}},
  \bibinfo{author}{\bibfnamefont{S.}~\bibnamefont{Yelin}},
  \bibinfo{author}{\bibfnamefont{M.~D.} \bibnamefont{Lukin}},
  \bibinfo{author}{\bibfnamefont{J.~I.} \bibnamefont{Cirac}}, \bibnamefont{and}
  \bibinfo{author}{\bibfnamefont{G.}~\bibnamefont{Giedke}},
  \bibinfo{journal}{Physical Review Letters} \textbf{\bibinfo{volume}{104}},
  \bibinfo{pages}{143601} (\bibinfo{year}{2010}).

\bibitem[{\citenamefont{Holstein and Primakoff}(1940)}]{Holstein1940}
\bibinfo{author}{\bibfnamefont{T.}~\bibnamefont{Holstein}} \bibnamefont{and}
  \bibinfo{author}{\bibfnamefont{H.}~\bibnamefont{Primakoff}},
  \bibinfo{journal}{Physical Review} \textbf{\bibinfo{volume}{58}},
  \bibinfo{pages}{1098} (\bibinfo{year}{1940}).

\bibitem[{\citenamefont{Kitagawa and Ueda}(1993)}]{Kitagawa:1993di}
\bibinfo{author}{\bibfnamefont{M.}~\bibnamefont{Kitagawa}} \bibnamefont{and}
  \bibinfo{author}{\bibfnamefont{M.}~\bibnamefont{Ueda}},
  \bibinfo{journal}{Physical Review A} \textbf{\bibinfo{volume}{47}},
  \bibinfo{pages}{5138} (\bibinfo{year}{1993}).

\bibitem[{\citenamefont{Urbaszek et~al.}(2012)\citenamefont{Urbaszek, Marie,
  Amand, Krebs, Voisin, Maletinsky, Hogele, and Imamoglu}}]{Urbaszek2012}
\bibinfo{author}{\bibfnamefont{B.}~\bibnamefont{Urbaszek}},
  \bibinfo{author}{\bibfnamefont{X.}~\bibnamefont{Marie}},
  \bibinfo{author}{\bibfnamefont{T.}~\bibnamefont{Amand}},
  \bibinfo{author}{\bibfnamefont{O.}~\bibnamefont{Krebs}},
  \bibinfo{author}{\bibfnamefont{P.}~\bibnamefont{Voisin}},
  \bibinfo{author}{\bibfnamefont{P.}~\bibnamefont{Maletinsky}},
  \bibinfo{author}{\bibfnamefont{A.}~\bibnamefont{Hogele}}, \bibnamefont{and}
  \bibinfo{author}{\bibfnamefont{A.}~\bibnamefont{Imamoglu}},
  \bibinfo{journal}{arXiv:1202.4637}  (\bibinfo{year}{2012}).

\bibitem[{\citenamefont{Korbicz et~al.}(2005)\citenamefont{Korbicz, Cirac, and
  Lewenstein}}]{Korbicz2005}
\bibinfo{author}{\bibfnamefont{J.~K.} \bibnamefont{Korbicz}},
  \bibinfo{author}{\bibfnamefont{J.~I.} \bibnamefont{Cirac}}, \bibnamefont{and}
  \bibinfo{author}{\bibfnamefont{M.}~\bibnamefont{Lewenstein}},
  \bibinfo{journal}{Physical Review Letters} \textbf{\bibinfo{volume}{95}},
  \bibinfo{pages}{120502} (\bibinfo{year}{2005}).

\bibitem[{\citenamefont{Pezz{\'e} and Smerzi}(2009)}]{Pezze2009}
\bibinfo{author}{\bibfnamefont{L.}~\bibnamefont{Pezz{\'e}}} \bibnamefont{and}
  \bibinfo{author}{\bibfnamefont{A.}~\bibnamefont{Smerzi}},
  \bibinfo{journal}{Physical Review Letters} \textbf{\bibinfo{volume}{102}},
  \bibinfo{pages}{100401} (\bibinfo{year}{2009}).

\bibitem[{\citenamefont{Hyllus et~al.}(2012)\citenamefont{Hyllus, Laskowski,
  Krischek, Schwemmer, Wieczorek, Weinfurter, Pezz{\'e}, and
  Smerzi}}]{Hyllus2012}
\bibinfo{author}{\bibfnamefont{P.}~\bibnamefont{Hyllus}},
  \bibinfo{author}{\bibfnamefont{W.}~\bibnamefont{Laskowski}},
  \bibinfo{author}{\bibfnamefont{R.}~\bibnamefont{Krischek}},
  \bibinfo{author}{\bibfnamefont{C.}~\bibnamefont{Schwemmer}},
  \bibinfo{author}{\bibfnamefont{W.}~\bibnamefont{Wieczorek}},
  \bibinfo{author}{\bibfnamefont{H.}~\bibnamefont{Weinfurter}},
  \bibinfo{author}{\bibfnamefont{L.}~\bibnamefont{Pezz{\'e}}},
  \bibnamefont{and} \bibinfo{author}{\bibfnamefont{A.}~\bibnamefont{Smerzi}},
  \bibinfo{journal}{Physical Review A} \textbf{\bibinfo{volume}{85}},
  \bibinfo{pages}{022321} (\bibinfo{year}{2012}).

\bibitem[{\citenamefont{Clifford and Clifford}(1999)}]{Clifford:1999uv}
\bibinfo{author}{\bibfnamefont{A.}~\bibnamefont{Clifford}} \bibnamefont{and}
  \bibinfo{author}{\bibfnamefont{T.}~\bibnamefont{Clifford}},
  \emph{\bibinfo{title}{{Fundamentals of Supercritical Fluids}}}
  (\bibinfo{publisher}{Oxford University Press}, \bibinfo{year}{1999}).

\bibitem[{\citenamefont{Bowden and Sung}(1979)}]{Bowden1979}
\bibinfo{author}{\bibfnamefont{C. M.}~\bibnamefont{Bowden}} \bibnamefont{and}
  \bibinfo{author}{\bibfnamefont{C. C.}~\bibnamefont{Sung}},
  \bibinfo{journal}{Physical Review A} \textbf{\bibinfo{volume}{19}},
  \bibinfo{pages}{2392} (\bibinfo{year}{1979}).

\bibitem[{\citenamefont{Kessler}()}]{Kessler:QGYw0mMI}
\bibinfo{author}{\bibfnamefont{E.~M.} \bibnamefont{Kessler}},
  \bibinfo{journal}{to be published}  (2012).

\bibitem[{\citenamefont{Lax}(1963)}]{Lax:1963cy}
\bibinfo{author}{\bibfnamefont{M.}~\bibnamefont{Lax}},
  \bibinfo{journal}{Physical Review} \textbf{\bibinfo{volume}{129}},
  \bibinfo{pages}{2342} (\bibinfo{year}{1963}).

\bibitem[{\citenamefont{Kim et~al.}(2002)\citenamefont{Kim, Lee, and
  Munro}}]{Kim2002}
\bibinfo{author}{\bibfnamefont{M. S.}~\bibnamefont{Kim}},
  \bibinfo{author}{\bibfnamefont{J.}~\bibnamefont{Lee}}, \bibnamefont{and}
  \bibinfo{author}{\bibfnamefont{W. J.}~\bibnamefont{Munro}},
  \bibinfo{journal}{Physical Review A} \textbf{\bibinfo{volume}{66}},
  \bibinfo{pages}{030301(R)} (\bibinfo{year}{2002}).

\bibitem[{\citenamefont{Nagy et~al.}(2010)\citenamefont{Nagy, Konya, Szirmai,
  and Domokos}}]{Nagy2010}
\bibinfo{author}{\bibfnamefont{D.}~\bibnamefont{Nagy}},
  \bibinfo{author}{\bibfnamefont{G.}~\bibnamefont{Konya}},
  \bibinfo{author}{\bibfnamefont{G.}~\bibnamefont{Szirmai}}, \bibnamefont{and}
  \bibinfo{author}{\bibfnamefont{P.}~\bibnamefont{Domokos}},
  \bibinfo{journal}{Physical Review Letters} \textbf{\bibinfo{volume}{104}},
  \bibinfo{pages}{130401} (\bibinfo{year}{2010}).

\bibitem[{\citenamefont{Lambert et~al.}(2004)\citenamefont{Lambert, Emary, and
  Brandes}}]{Lambert04}
\bibinfo{author}{\bibfnamefont{N.}~\bibnamefont{Lambert}},
  \bibinfo{author}{\bibfnamefont{C.}~\bibnamefont{Emary}}, \bibnamefont{and}
  \bibinfo{author}{\bibfnamefont{T.}~\bibnamefont{Brandes}},
  \bibinfo{journal}{Phys. Rev. Lett.} \textbf{\bibinfo{volume}{92}},
  \bibinfo{pages}{073602} (\bibinfo{year}{2004}).

\bibitem[{\citenamefont{Drummond and Walls}(1980)}]{Drummond1980}
\bibinfo{author}{\bibfnamefont{P.~D.} \bibnamefont{Drummond}} \bibnamefont{and}
  \bibinfo{author}{\bibfnamefont{D.~F.} \bibnamefont{Walls}},
  \bibinfo{journal}{Journal of Physics A: Mathematical and General}
  \textbf{\bibinfo{volume}{13}}, \bibinfo{pages}{725} (\bibinfo{year}{1980}).

\bibitem[{\citenamefont{Jelezko and Wrachtrup}(2006)}]{Jelezko:2006}
\bibinfo{author}{\bibfnamefont{F.}~\bibnamefont{Jelezko}} \bibnamefont{and}
  \bibinfo{author}{\bibfnamefont{J.}~\bibnamefont{Wrachtrup}},
  \bibinfo{journal}{physica status solidi (a)} \textbf{\bibinfo{volume}{203}},
  \bibinfo{pages}{3207} (\bibinfo{year}{2006}).

\bibitem[{\citenamefont{Togan et~al.}(2011)\citenamefont{Togan, Chu, Imamoglu,
  and Lukin}}]{Togan:2011}
\bibinfo{author}{\bibfnamefont{E.}~\bibnamefont{Togan}},
  \bibinfo{author}{\bibfnamefont{Y.}~\bibnamefont{Chu}},
  \bibinfo{author}{\bibfnamefont{A.}~\bibnamefont{Imamoglu}}, \bibnamefont{and}
  \bibinfo{author}{\bibfnamefont{M.~D.} \bibnamefont{Lukin}},
  \bibinfo{journal}{Nature} \textbf{\bibinfo{volume}{478}},
  \bibinfo{pages}{497} (\bibinfo{year}{2011}).

\bibitem[{\citenamefont{Chisholm et~al.}(2011)\citenamefont{Chisholm, Maurer,
  Kucsko, Lo, Yao, Shields, Park, and Lukin}}]{Chisholm:2011uj}
\bibinfo{author}{\bibfnamefont{N.}~\bibnamefont{Chisholm}},
  \bibinfo{author}{\bibfnamefont{P.}~\bibnamefont{Maurer}},
  \bibinfo{author}{\bibfnamefont{G.}~\bibnamefont{Kucsko}},
  \bibinfo{author}{\bibfnamefont{P.}~\bibnamefont{Lo}},
  \bibinfo{author}{\bibfnamefont{N.}~\bibnamefont{Yao}},
  \bibinfo{author}{\bibfnamefont{B.}~\bibnamefont{Shields}},
  \bibinfo{author}{\bibfnamefont{H.}~\bibnamefont{Park}}, \bibnamefont{and}
  \bibinfo{author}{\bibfnamefont{M.}~\bibnamefont{Lukin}},
  \bibinfo{journal}{American Physical Society, 42nd Annual Meeting of the APS Division of Atomic, Molecular and Optical Physics, abstract \#OPE.10} 
  (\bibinfo{year}{2011}).

\bibitem[{\citenamefont{Kroner et~al.}(2008)\citenamefont{Kroner, Weiss,
  Biedermann, Seidl, Manus, Holleitner, Badolato, Petroff, Gerardot, Warburton
  et~al.}}]{Kroner:2008kf}
\bibinfo{author}{\bibfnamefont{M.}~\bibnamefont{Kroner}},
  \bibinfo{author}{\bibfnamefont{K. M.}~\bibnamefont{Weiss}},
  \bibinfo{author}{\bibfnamefont{B.}~\bibnamefont{Biedermann}},
  \bibinfo{author}{\bibfnamefont{S.}~\bibnamefont{Seidl}},
  \bibinfo{author}{\bibfnamefont{S.}~\bibnamefont{Manus}},
  \bibinfo{author}{\bibfnamefont{A. W.}~\bibnamefont{Holleitner}},
  \bibinfo{author}{\bibfnamefont{A.}~\bibnamefont{Badolato}},
  \bibinfo{author}{\bibfnamefont{P. M.}~\bibnamefont{Petroff}},
  \bibinfo{author}{\bibfnamefont{B. D.}~\bibnamefont{Gerardot}},
  \bibinfo{author}{\bibfnamefont{R. J.}~\bibnamefont{Warburton}},
  \bibinfo{author}{\bibfnamefont{K.}~\bibnamefont{Karrai}}, \bibinfo{journal}{Physical Review Letters}
  \textbf{\bibinfo{volume}{100}},  \bibinfo{pages}{156803} (\bibinfo{year}{2008}).

\bibitem[{\citenamefont{Carmichael}(1999)}]{Carmichael1999}
\bibinfo{author}{\bibfnamefont{H.~J.} \bibnamefont{Carmichael}},
  \emph{\bibinfo{title}{{Statistical Methods in Quantum Optics 1}}}
  (\bibinfo{publisher}{Springer}, \bibinfo{address}{Berlin},
  \bibinfo{year}{1999}).

\end{thebibliography}
\end{document}